\documentclass[]{aa}  
\bibpunct{(}{)}{;}{a}{}{,}
\usepackage{graphicx}
\usepackage{txfonts}
\usepackage{color}
\usepackage{supertabular}

\begin{document}

\title{AT 2019abn: multi-wavelength observations over the first 200~days\thanks{Table~\ref{tab:phot} is available in electronic form
at the CDS via anonymous ftp to cdsarc.u-strasbg.fr (130.79.128.5)
or via http://cdsarc. u-strasbg.fr/viz-bin/cat/J/A+A/637/A20}}

\author{S.~C. Williams\inst{1,2,3},
D. Jones\inst{4,5},
P. Pessev\inst{6,4},
S. Geier\inst{6,4},
R.~L.~M. Corradi\inst{6,4},
I.~M. Hook\inst{1},
M.~J. Darnley\inst{7},
O. Pejcha\inst{8},\\
A. N\'u\~nez\inst{6,4},
S. Meingast\inst{9},
S. Moran\inst{3,10}
}
\institute{Physics Department, Lancaster University, Lancaster, LA1 4YB, UK\\
\email{steven.williams@utu.fi}
\and
Finnish Centre for Astronomy with ESO (FINCA), Quantum, Vesilinnantie 5, University of Turku, 20014 Turku, Finland
\and
Department of Physics and Astronomy, University of Turku, 20014 Turku, Finland
\and
Instituto de Astrof\'isica de Canarias, E-38205 La Laguna, Tenerife, Spain
\and
Departamento de Astrof\'isica, Universidad de La Laguna, E-38206 La Laguna, Tenerife, Spain
\and
Gran Telescopio Canarias (GRANTECAN), Cuesta de San Jos\'e s/n, 38712 Bre\~na Baja, La Palma, Spain
\and
Astrophysics Research Institute, Liverpool John Moores University, IC2 Liverpool Science Park, Liverpool, L3 5RF, UK
\and
Institute of Theoretical Physics, Faculty of Mathematics and Physics, Charles University, V~Hole\v{s}ovi\v{c}k\'{a}ch 2, 180 00, Praha 8, Czech Republic
\and
University of Vienna, T\"urkenschanzstrasse 17, 1180, Vienna, Austria
\and
Nordic Optical Telescope, Apartado 474, E-38700 Santa Cruz de La Palma, Spain
}

\authorrunning{S.~C. Williams et al.}
\titlerunning{AT~2019abn: the first 200 days}

\date{Received 9 Dec 2019; Accepted 18 Jan 2020}

 
  \abstract
    {AT~2019abn was discovered in the nearby M51 galaxy by the Zwicky Transient Facility at more than two magnitudes and around three\,weeks prior to its optical peak.}
    {We aim to conduct a detailed photometric and spectroscopic follow-up campaign for AT~2019abn, with early discovery allowing for significant pre-maximum observations of an intermediate luminosity red transient (ILRT) for the first time.}
    {This work is based on the analysis of $u'BVr'i'z'H$ photometry and low-resolution spectroscopy using the Liverpool Telescope, medium-resolution spectroscopy with the Gran Telescopio Canarias (GTC), and near-infrared imaging with the GTC and the Nordic Optical Telescope.}
   {We present the most detailed optical light curve of an ILRT to date, with multi-band photometry starting around three weeks before peak brightness. The transient peaked at an observed absolute magnitude of $M_{r'}=-13.1$, although it is subject to significant reddening from dust in M51, implying an intrinsic $M_{r'}\sim-15.2$. The initial light curve showed a linear, achromatic rise in magnitude before becoming bluer at peak. After peak brightness, the transient gradually cooled. This is reflected in our spectra, which at later times show absorption from such species as Fe~{\sc i}, Ni~{\sc i} and Li~{\sc i}. A spectrum taken around peak brightness shows narrow, low-velocity absorption lines, which we interpret as likely to originate from pre-existing circumstellar material.}
   {We conclude that while there are some peculiarities, such as the radius evolution, AT~2019abn fits in well overall with the ILRT class of objects and is the most luminous member of the class seen to date.}


   \maketitle
%

\section{Introduction}
In recent years, an increasing number of transients have been observed to peak with optical luminosities between those of novae (typically $M_V>-10$) and supernovae (SNe; typically $M_V<-15$). This is due in large part to the increasing coverage, depth, and cadence of astronomical surveys. Some of these intermediate luminosity objects are particularly red in colour and those can typically be placed in one of two categories:

\textit{Luminous red novae} (LRNe) -- The first well-observed LRN was V838 Monocerotis in our Galaxy in 2002 \citep{2002A&A...389L..51M,2005MNRAS.360.1281R}. The pre-outburst light curve of red nova V1309 Scorpii revealed it to be the result of a compact binary merger \citep{2011A&A...528A.114T}. A number of similar events have now been observed, including extragalactic examples in M31 (e.g.\ \citealp{2015A&A...578L..10K,2015ApJ...805L..18W,Williams_2019}) and M101 \citep{2015A&A...578L..10K,2017ApJ...834..107B}.

\textit{Intermediate luminosity red transients} (ILRTs) -- Several extragalactic examples of ILRTs have been observed over the last decade or so. See, for example, M85~OT~2006-1 \citep{2007Natur.447..458K}, SN~2008S \citep{2009MNRAS.398.1041B,2009ApJ...697L..49S}, NGC~300~OT~2008-1 (hereafter NGC~300~OT; \citealp{2009ApJ...699.1850B,2009ApJ...695L.154B}), and AT~2017be \citep{2018MNRAS.480.3424C}. These are generally more luminous than LRNe and, in some cases, they have been found to be associated with dusty progenitor stars \citep{2008ApJ...681L...9P,2009ApJ...699.1850B,2009ApJ...695L.154B}. SN~2008S and NGC~300~OT are also discussed in detail by \citet{2011ApJ...741...37K}.

The spectra of LRNe and ILRTs at peak are broadly similar, showing Balmer emission on a F-type supergiant-like spectrum. However, after peak, the two classes deviate, with the LRNe rapidly reddening and becoming cool enough for strong TiO absorption bands to appear in the spectrum \citep[e.g.][]{2005MNRAS.360.1281R,2015ApJ...805L..18W}. ILRTs have shown slower spectroscopic evolution and can typically be identified by strong and narrow [Ca~{\sc ii}] emission lines, although these have also been observed in LRNe \citep{2019arXiv190913147C}. To unambiguously distinguish between the two categories of object requires the transients to be followed over the course of several months in some cases.

The nature of ILRTs has not been settled, with plausible scenarios including, for example, an electron capture SN (see e.g. \citealp{2009MNRAS.398.1041B}) or a giant eruption from a luminous blue variable (LBV; see e.g. \citealp{2009ApJ...697L..49S}). Both SN~2008S and NGC~300~OT eventually became fainter than their respective progenitors \citep{2016MNRAS.460.1645A}. \citet{2016MNRAS.460.1645A} suggest that this may point to a weak SN scenario, as extreme dust models were needed to reconcile the very late-time observations with a surviving star. \citet{2010MNRAS.403..474W} modelled the pre-existing dust around SN~2008S as amorphous carbon grains and indeed found that the observations were inconsistent with silicate grains making up a significant component of the dust. See \citet{2019NatAs...3..676P} for a recent review of the different classes of intermediate luminosity transients.

AT 2019abn (ZTF19aadyppr) was first detected as a transient in M51 by the Zwicky Transient Facility (ZTF; \citealp{2019PASP..131a8002B}) on 2019 Jan 22.56~UT and announced following the second detection on 2019 Jan 25.51, at $13^{\mathrm{h}}29^{\mathrm{m}}42^{\mathrm{s}}\!.394$ $+47^{\circ}11^{\prime}16^{\prime\prime}\!\!.99$ (J2000; \citealp{2019TNSTR.141....1N}), by \texttt{AMPEL} (Alert Management, Photometry and Evaluation of Lightcurves; \citealp{2019arXiv190405922N}). In this work we refer to this first optical detection on 2019 Jan 22.56~UT as the discovery date ($t=0$).

\citet{2019arXiv190407857J} presented the discovery of this object and its variable progenitor system, along with its evolution over the first $\sim$80 days after discovery. Here we present detailed photometric and spectroscopic observations of the first $\sim$200 days after discovery, including the best observed early light curve of any ILRT to date. Our photometry begins 3.7\,days after discovery (0.7\,d after it was announced) and our spectra span the range from 7.7 to 165.4\,days after discovery.

\section{Observations and data reduction}
\subsection{Liverpool Telescope photometry}
We obtained multi-colour follow-up using the IO:O \citep{smith_steele_2017} and IO:I \citep{2016JATIS...2a5002B} imagers on the 2\,m Liverpool Telescope (LT; \citealp{2004SPIE.5489..679S}). We used SDSS $u'r'i'z'$ and Bessel \textit{BV} filters in IO:O and \textit{H}-band imaging with IO:I. AT~2019abn is coincident with significant dust absorption in M51, meaning the background at the position of the transient is not captured well by an annulus and template subtraction is required. For the $u'r'i'z'$ observations, we used archival Sloan Digital Sky Survey (SDSS) DR12 \citep{2015ApJS..219...12A} images for template subtraction and for the \textit{BV} observations, we used archival LT observations. The template subtraction was performed using standard routines in \texttt{IRAF}\footnote{IRAF is distributed by the National Optical Astronomy Observatory, which is operated by the Association of Universities for Research in Astronomy (AURA) under a cooperative agreement with the National Science Foundation.} \citep{1986SPIE..627..733T,1993ASPC...52..173T}. Template subtraction was not performed on the \textit{H}-band images, however the effect in this band is expected to be relatively small as the interstellar dust absorption will be lower at these wavelengths.

As M51 takes up a large fraction of the IO:I imager's field of view (FoV), the dithering of the telescope did not result in a good sky subtraction. We therefore obtained offset\footnote{The offset sky observations were centred at the position $13^{\mathrm{h}}28^{\mathrm{m}}24^{\mathrm{s}}\!.66$ $+47^{\circ}05^{\prime}10^{\prime\prime}\!\!.8$ (J2000).} sky frames immediately after each \textit{H}-band observation (except for the first observation). These offset sky frames were combined, then scaled to and subtracted from each individual science frame. The sky-subtracted frames were then aligned and combined to produce the final \textit{H}-band images on which photometry was performed.

\subsection{Nordic Optical Telescope photometry}
AT~2019abn was observed with the near-infrared instrument NOTCam at the 2.56\,m Nordic Optical Telescope (NOT) at 2019 Jun~13.94\,UT in the $K_{\mathrm{s}}$-band and at 2019 Jul 21.89 in the \textit{J}, \textit{H}, and $K_{\mathrm{s}}$-bands. As the angular extension of M51 on the sky is significantly bigger than the FoV of NOTCam, beam-switching with an sky-offset of 5\,arcmin was used to allow for a good sky-subtraction. A 9-point dither pattern was used with $5\times4$\,s ramp-sampling exposures, resulting in a total on-source time of 180\,s in each filter, and the same time spent on the sky observation. The images were reduced using a custom pipeline written in \texttt{IDL}. For sky-subtraction, a median combination of the sky exposures was used which was then scaled to the background level of the object exposures. The sky-subtracted images were then aligned by the WCS coordinates and mean-combined. Bad pixels were treated by discarding them in the combination.

\subsection{Gran Telescopio Canarias photometry}
Near-infrared images of the transient were also obtained at the 10.4\,m Gran Telescopio Canarias (GTC) with the EMIR instrument \citep{2007hsa..conf...81G} as part of an outreach programme. AT~2019abn was observed at 2019 April 12.97\,UT through the $JHK_{\mathrm{s}}$ 2MASS filters under grey sky (moon at 51\% illumination), 1.1$^{\prime\prime}$ seeing and variable atmospheric transparency. In each filter, a series of exposures were taken, adopting a seven point dithering pattern with 10$^{\prime\prime}$ offsets. Since the host galaxy covers a large part of the EMIR imaging mode FoV ($6.67\times6.67$\,arcmin), sky frames with a telescope offset of 7\,arcmin to the east were also obtained. Total exposure times were 280, 140, and 84\,s on source in the \textit{J}, \textit{H} and $K_{\mathrm{s}}$ filters, with individual frames duration of 10, 5, and 3\,s, respectively. The dark current of the EMIR detector is below 0.15 electrons\,s$^{-1}$, so the individual frames were directly divided by the corresponding twilight sky flats and stacked together in the final image using the Cambridge Astronomy Survey Unit \texttt{imstack} routine.

\subsection{Photometric Calibration}
Optimal photometry was performed on the processed images, except on a few occasions when there were tracking or guiding issues with the observations, which caused elongated and irregular PSFs. For these images, aperture photometry was used. All optimal and aperture photometry was performed using the \texttt{STARLINK} package \texttt{GAIA}\footnote{\url{http://star-www.dur.ac.uk/~pdraper/gaia/gaia.html}} \citep{2014ASPC..485..391C}.

All \textit{u$'$BVr$'$i$'$z$'$} photometry presented in this work was calibrated against SDSS DR9 \citep{2012ApJS..203...21A}. Due to the lack of stars with SDSS photometry in our FoV, we used the star J132918.06+471616.9, which has a magnitude of $u=19.137\pm0.023$,      $g=18.297\pm0.007$,     $r=17.902\pm0.007$, $i=17.741\pm0.007,$ and  $z=17.695\pm0.016$\,mag \citep{2012ApJS..203...21A}. The star also has very similar magnitude measurements in Pan-STARRS DR1 \citep{2016arXiv161205560C}. From the magnitude of this star, we calibrated a sequence of eight further standard stars in the FoV across several nights of observations. Each observation of AT~2019abn was then calibrated against the calculated magnitudes of these stars. The \textit{B} and \textit{V}-band magnitudes of the standard stars were computed using the transformations from \citet{2006A&A...460..339J} and then used to calibrate the \textit{BV} photometry of AT~2019abn. All \textit{JHK} photometry was calibrated against several stars from 2MASS \citep{2006AJ....131.1163S}.  Template subtraction was not performed on any of our $J$, $H,$ or $K_{\mathrm{s}}$-band data.

\subsection{Gran Telescopio Canarias spectroscopy}
AT~2019abn was observed on 2019~Feb~25.27, Apr~10.93, May~31.04, and Jul~6.92\,UT with the Optical System for Imaging and low-Intermediate-Resolution Integrated Spectroscopy (OSIRIS) instrument mounted on GTC as part of a Director's Discretionary Time award.  On each night, exposures of 945\,s were taken through a 0.6\arcsec{} wide long slit at the parallactic angle, first with the R2500V grating (4500\,\AA~$<\lambda<$~6000\,\AA), then the R2500R grating (5600\,\AA~$<\lambda<$~7600\,\AA), both of which provide a resolving power, $R\sim2500$.  The resulting spectra were then debiased, wavelength calibrated against HgAr, Xe and Ne arc lamp spectra, sky-subtracted, and optimally-extracted using the algorithm of \citet{1986PASP...98..609H}. 

Flux calibration was performed using observations of the standard star Ross 640 \citep{1974ApJS...27...21O}, taken on the same nights using the same instrumental set-up and reduced in the same manner. We performed an absolute flux calibration using our LT \textit{V} and $r'$-band photometry. These spectra were linearly warped to match the two photometry bands. This relative correction was generally relatively small, however, with the initial relative flux calibrations agreeing well across the two filters.

\subsection{Liverpool Telescope spectroscopy}
We obtained eight low-resolution (R~$\sim$~350) spectra with the SPectrograph for the Rapid Acquisition of Transients (SPRAT) on the LT. The spectra were reduced using the SPRAT pipeline \citep{2014SPIE.9147E..8HP} and an absolute flux calibration made using the \textit{V} and $r'$-band light curves in the same way as for the GTC spectra.  A log of our spectra of AT~2019abn is shown in Table~\ref{tab:log}. Line identification for our spectra was aided by the multiplet tables of \citet{1945CoPri..20....1M} and the NIST Atomic Spectra Database \citep{NIST}.

\begin{table}
\caption{Log of Gran Telescopio Canarias and Liverpool Telescope spectra taken of AT~2019abn.} \label{tab:log}      
\centering
\begin{tabular}{l c c c}          
\hline\hline                        
Instrument &Date [UT] &$t$ [days] &Resolution\\
\hline                                   
LT SPRAT &2019 Jan 30.21 &7.7 &350\\
LT SPRAT &2019 Feb 03.22 &11.7 &350\\
LT SPRAT &2019 Feb 06.23 &14.7 &350\\
LT SPRAT &2019 Feb 10.11 &18.6 &350\\
LT SPRAT &2019 Feb 11.11 &19.6 &350\\
LT SPRAT &2019 Feb 20.25 &28.7 &350\\
LT SPRAT &2019 Feb 24.07 &32.5 &350\\
GTC OSIRIS &2019 Feb 25.27 &33.7 &2500\\
GTC OSIRIS &2019 Apr 10.93 &78.4 &2500\\
LT SPRAT &2019 Apr 15.96 &83.4 &350\\
GTC OSIRIS &2019 May 31.04 &128.5 &2500\\
GTC OSIRIS &2019 Jul 06.92 &165.4 &2500\\
\hline       
\end{tabular}
\end{table}

\section{Photometric evolution}
\begin{figure}
\centering
\includegraphics[width=\columnwidth]{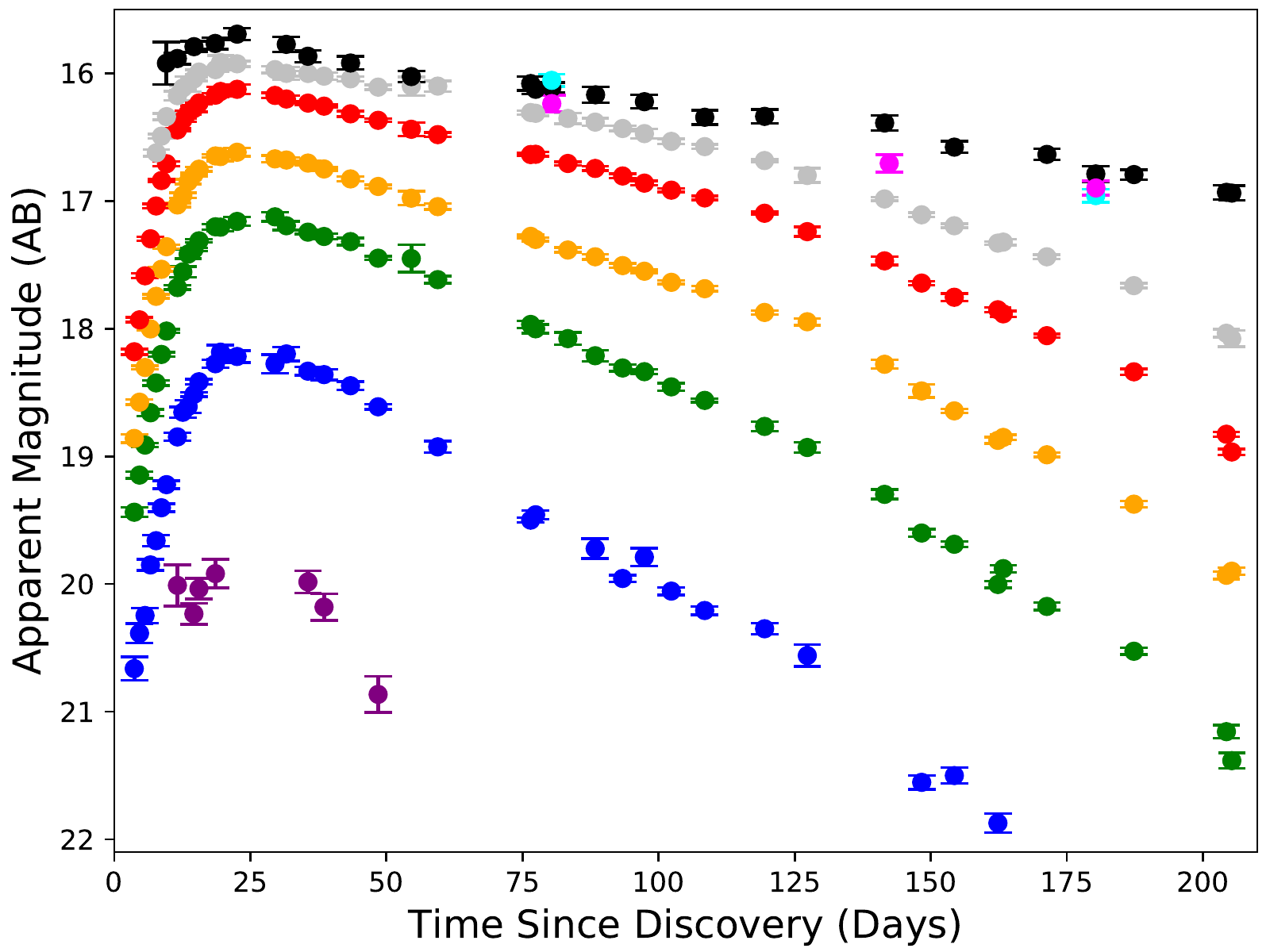}
\includegraphics[width=\columnwidth]{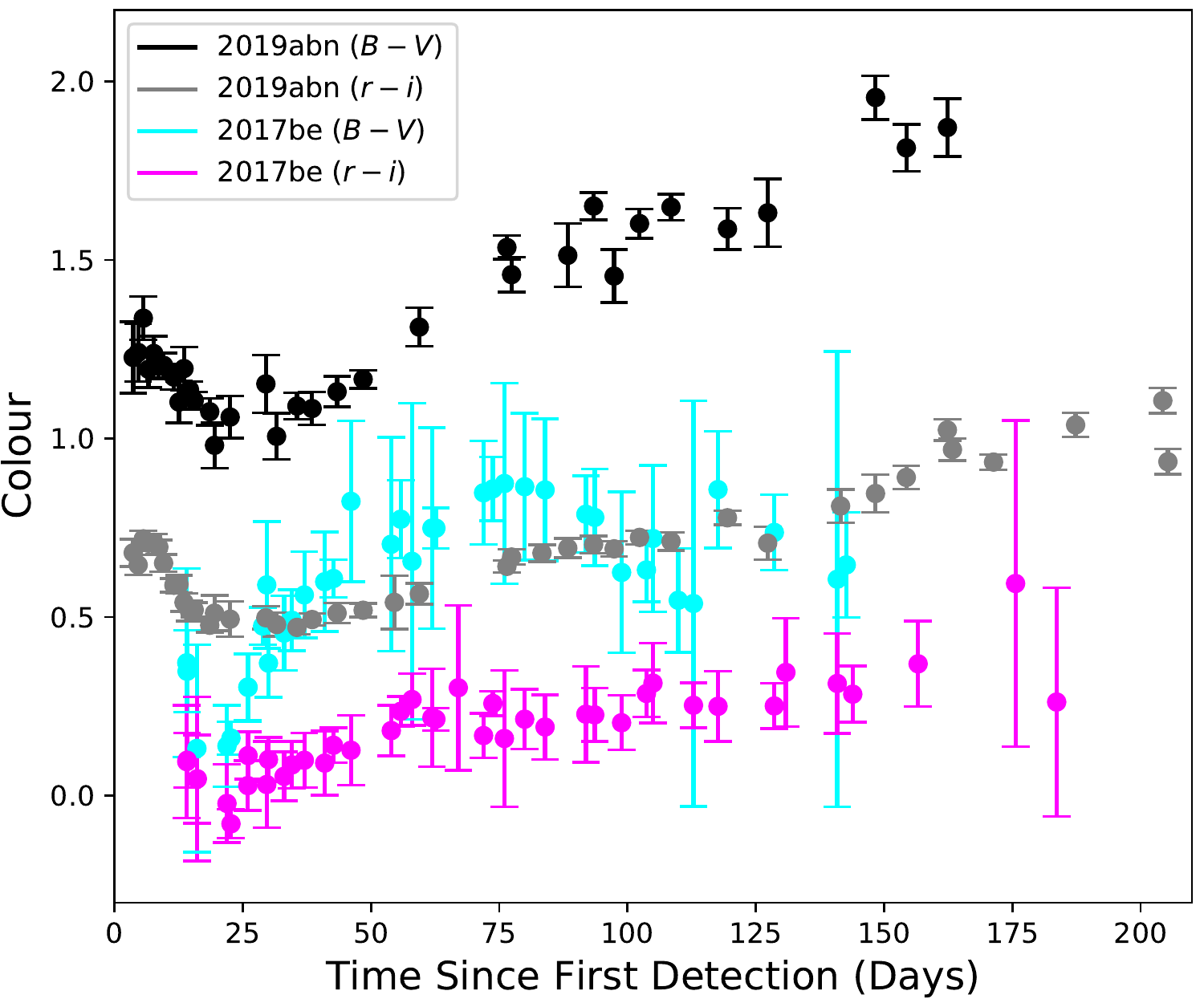}
\caption{\textit{Top:} Multi-colour light curve of AT~2019abn, with: $u'$, purple; \textit{B}, blue; \textit{V}, green; $r'$, orange; $i'$, red; $z'$, grey; \textit{H}, black (they appear on the light curve from faintest to brightest in that order). The \textit{J}-band photometry is in cyan, with $K_{\mathrm{s}}$-band in magenta. \textit{Bottom:} $(B-V)$ and $(r'-i')$ colour evolution (again in AB mags) of AT~2019abn, compared to that of AT~2017be. The photometry for both objects is corrected for Galactic reddening only ($E_{B-V}=0.03$ and $E_{B-V}=0.05$\,mag respectively), and Vega \textit{BV} mags are converted to the AB system using \citet{1998A&A...333..231B}.}
\label{fig:lc}
\end{figure}

Our typically daily $BVr'i'$ photometry of AT~2019abn during the rise to peak optical brightness yields the best early-time light curve of any ILRT to date, beginning $>$2\,mag prior to peak in each filter. The full light curve of AT~2019abn is shown in Figure~\ref{fig:lc}. We find that both the first stage of the decline, and the early rise are well described by linear (in magnitude vs time) fits. Here we refer to the initial rise as before the light curve begins to turnover as it approaches peak ($t<9$\,d; our first six data points). These linear rises and declines are summarised in Table~\ref{tab:decl}. The initial rise rate has no strong colour dependency and generally at the level of $\sim$0.26 mag\,day$^{-1}$.

\begin{table*}
\caption{Light-curve parameters of AT~2019abn in each band.  Rise times are computed based on the first six data points ($t<9$\,d), before the light curve starts to turnover as it gets closer to peak. The plateau decline rates are fit between 25 and 130 days after discovery.} \label{tab:decl}      
\centering
\begin{tabular}{c c c c c c}          
\hline\hline                        
Filter & Rise rate [mag\,day$^{-1}$] &Decline rate [mag\,day$^{-1}$] &Peak magnitude [AB] &Time of peak [MJD] &Peak - $t_0$ [days]\\
\hline                                   
\textit{B} &$0.252\pm0.013$ &$0.0264\pm0.0008$ &$18.17\pm0.04$ &$58530.4\pm2.1$ &$24.9\pm2.1$\\
\textit{V} &$0.240\pm0.005$ &$0.0184\pm0.0002$ &$17.12\pm0.02$ &$58530.4\pm2.0$ &$24.9\pm2.0$\\  
$r'$ &$0.265\pm0.009$ &$0.0136\pm0.0001$ &$16.59\pm0.02$ &$58529.1\pm1.0$ & $23.5\pm1.0$\\
$i'$ &$0.268\pm0.011$ &$0.0102\pm0.0001$ &$16.12\pm0.02$ &$58528.6\pm2.0$ &$23.0\pm1.0$\\
$z'$ &... &$0.0081\pm0.0002$ &$15.93\pm0.02$ &$58526.9\pm1.9$ &$21.4\pm1.9$\\
\textit{H} &... &$0.0060\pm0.0004$ &$15.73\pm0.04$ &$58527.6\pm2.4$ &$22.1\pm2.4$\\
\hline       
\end{tabular}
\end{table*}

To derive the date and magnitude of peak brightness for each filter, we fitted a cubic spline to the light curve data. The exact times of peak brightness in each filter, shown in Table~\ref{tab:decl}, have relatively large uncertainties. This is due to the portion of the light curve around peak being approximately flat for a number of days; therefore, even when they are relatively small, the errors on the photometry around this time lead to a much larger uncertainty on the date of maximum. The peak magnitudes themselves are well constrained however. Given the complex field, the systematic errors on the photometry are likely larger than the (statistical) errors derived from the fitting (only the latter of which are shown in Table~\ref{tab:decl}).

Taking the distance modulus of M51 to be $29.67\pm0.02$\,mag ($8.58\pm0.10$\,Mpc; \citealp{2016ApJ...826...21M}), and correcting for the small amount of foreground Galactic extinction (see Section~\ref{ext}), we find a peak absolute magnitude of $M_{r'}=-13.08\pm0.04$ (statistical errors only). However, the transient is subject to substantial additional reddening internal to M51, which we estimate to be between $E_{B-V}=0.79$ and 0.9 (taking $R_V=3.1$; see Section~\ref{ext}). This implies that the intrinsic absolute peak is $M_{r'}=-15.2\pm0.2$, making AT~2019abn the most luminous ILRT to date.

\begin{figure}
\centering
\includegraphics[width=\columnwidth]{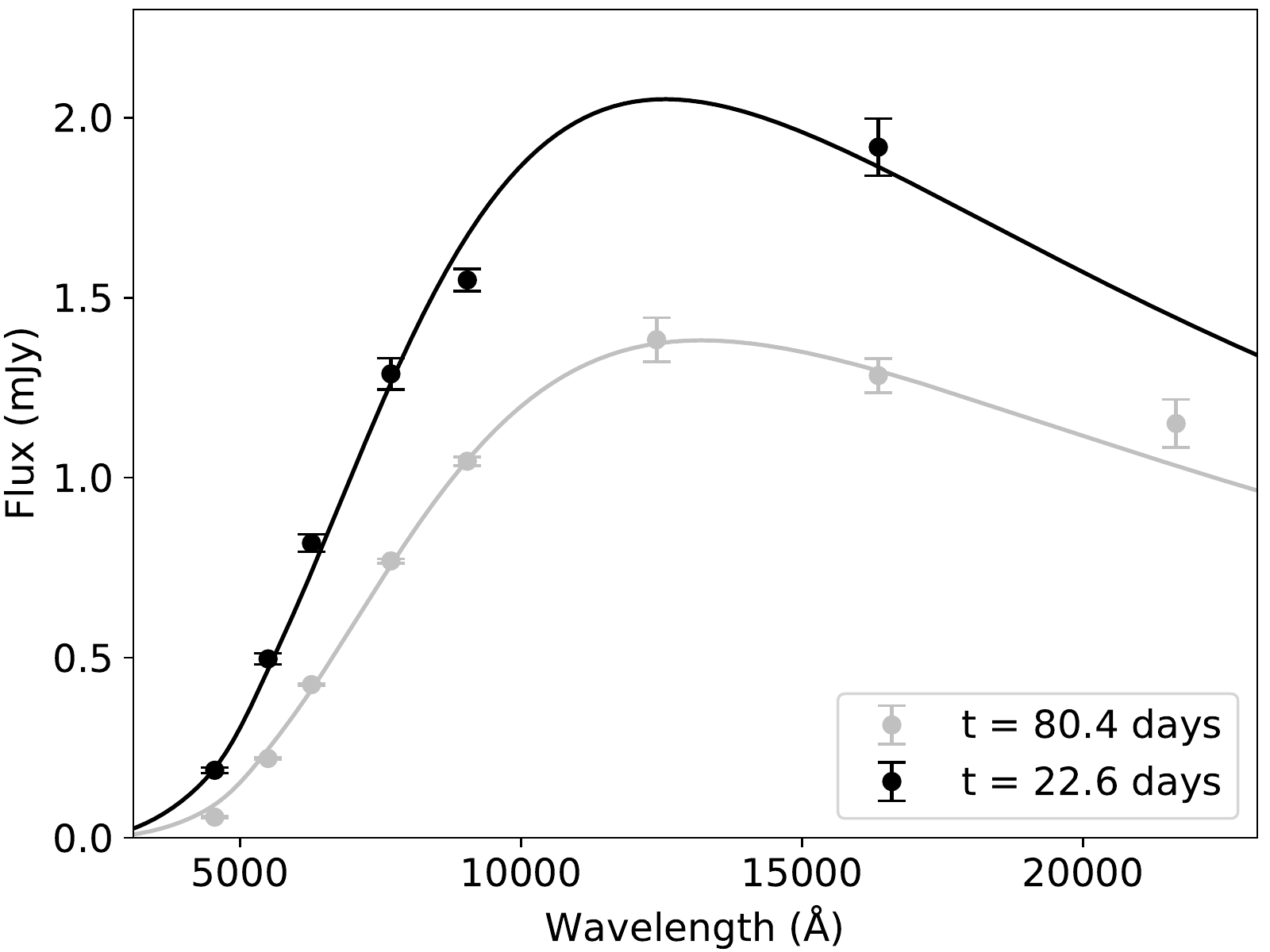}
\includegraphics[width=\columnwidth]{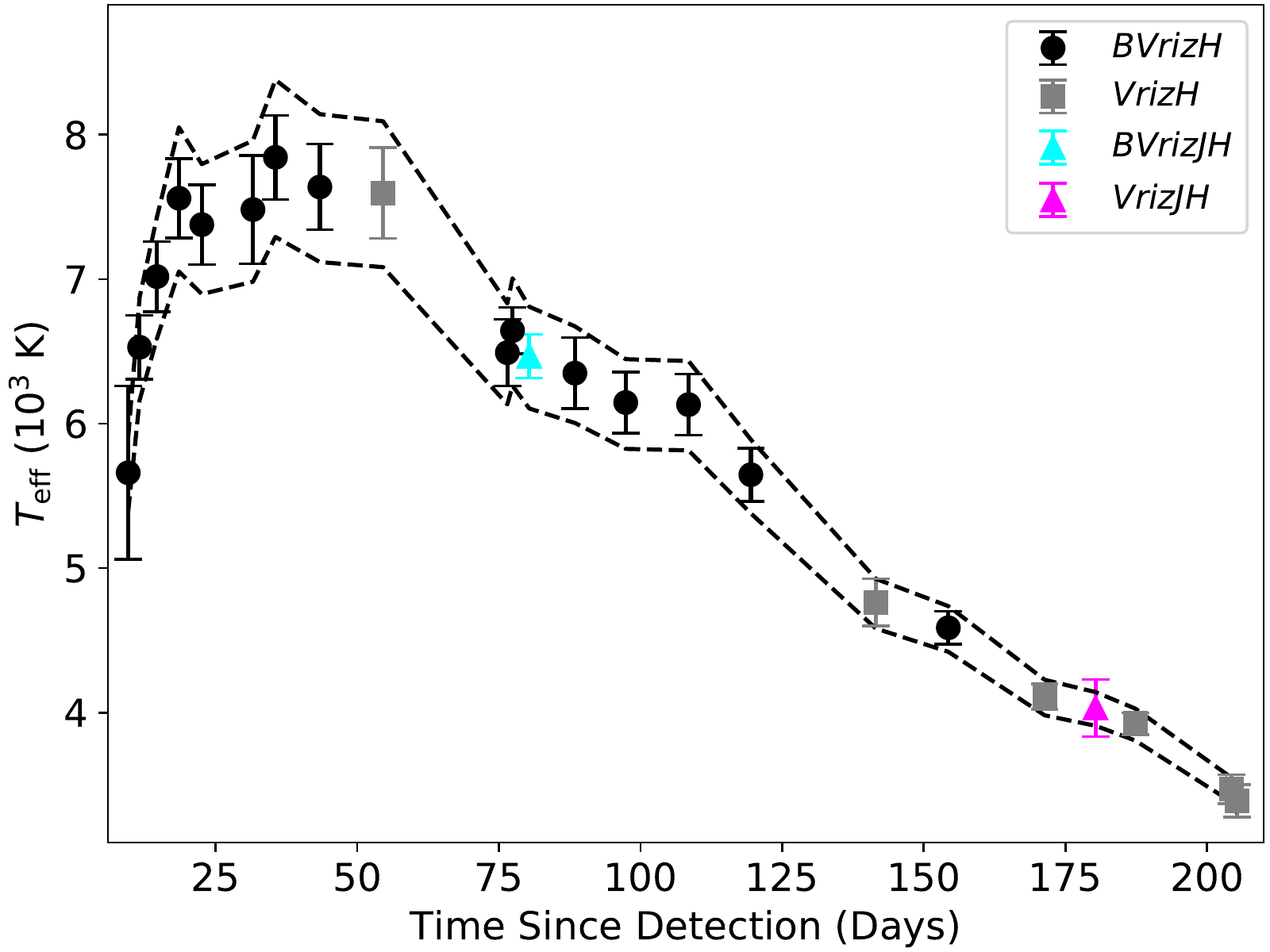}
\caption{\textit{Top:} Two example SEDs and the best-fit reddened blackbody function. The $K_{\mathrm{s}}$ photometry is shown for the $t=80.4$\,d epoch, although is not included in the fitting. \textit{Bottom:} Temperature evolution of AT~2019abn, as derived from SED fitting of the optical-NIR photometry and assuming $E_{B-V}=0.85$. \textit{B}-band was used in the fitting when available. All $BVr'i'z'H$ epochs shown in the plot were also fitted using $Vr'i'z'H$, which agreed well. For clarity we therefore only show the $Vr'i'z'H$ fits at epochs with no \textit{B}-band. The filters included in the fitting of each point are shown in the legend. The dashed lines illustrate the evolution for $E_{B-V}=0.90$ (upper) and $E_{B-V}=0.79$ (lower).}
\label{fig:temp}
\end{figure}

In the $u'$-band we see a rapid decline after peak. Between $t=35.6$ and 48.5\,days, AT~2019abn fades by $0.88\pm0.17$\,mag in the $u'$-band, while fading by only $0.28\pm0.04$\,mag in the \textit{B}-band. Given we see the appearance of many metal absorption lines between the two GTC spectra taken at $t=33.7$\,d and 78.4\,d, we interpret this $u'$-band drop-off as being predominately caused by increased line blanketing as the temperature falls. This rapid change in $(u'-B)$ colour is inconsistent with what is expected from a blackbody given the temperature evolution at this time (see Figure~\ref{fig:temp}). We note that ILRT  AT~2017be also displayed a rapid decline in the \textit{u}-band \citep{2018MNRAS.480.3424C}.

There appears to be a break in the light curve after around 130\,days, where the transient begins to decline more rapidly. This is not seen in the \textit{H}-band, where the later decline is consistent with the linear fit to the earlier data, and there is too little data in the \textit{B}-band to be able to see any change. We find later decline rates of $0.0281\pm0.0017$ (\textit{V}), $0.0247\pm0.0009$ ($r'$), $0.0221\pm0.0011$ ($i'$) and $0.0156\pm0.0005$\,mag\,day$^{-1}$ ($z'$).

The overall relative colour evolution of AT~2019abn and AT~2017be (Figure~\ref{fig:lc}, lower panel) is broadly similar, although the timescales appear different. This may be expected given the light curves of ILRTs vary considerably in the evolution timescale (see Section~\ref{sec:com} for further discussion; also see e.g. Figure~4 of \citealp{2018MNRAS.480.3424C}). What is clear from this comparison is that AT~2019abn has much redder apparent colours than AT~2017be.

To estimate the temperature evolution of AT~2019abn, we first correct for the well-constrained Galactic reddening of $E_{B-V}=0.03$ (see Section~\ref{ext}) assuming $R_V=3.1$ and a \citet{1999PASP..111...63F} reddening law. We then use the additional (i.e.\ M51 interstellar + circumstellar) extinction derived in Section~\ref{ext} to fit a reddened blackbody function to our photometry. Given our peak temperature is tied to the initial assumption we use to estimate the extinction at peak ($7000\leq{T_{\mathrm{eff}}}\leq8000$\,K), we can only interpret the relative change in temperature over the evolution of AT~2019abn and the precise value, and uncertainty, of the peak temperature is largely meaningless. 

We fit our photometry (\textit{B}-band to \textit{H}-band) using a reddened blackbody function, with $E_{B-V}=0.85$ and $R_V=3.1$, as derived in Section~\ref{ext}. Where possible we use all filters between these two bands in the fitting, however, we only have two \textit{J}-band observations, so this is normally not included. The $K_{\mathrm{s}}$-band observations are not included in the fitting as they may be contaminated by re-emission from circumstellar dust. At some epochs there is no \textit{B}-band detection, particularly at late times as AT~2019abn fades. In these cases we fit the SED using \textit{V}-band to \textit{H}-band. For epochs with \textit{B}-band to \textit{H}-band data, we also fit the SED using just \textit{V}-band to \textit{H}-band observations, and found very little difference in the implied temperature. An example of the blackbody fits to two epochs of photometry is shown in the upper panel of Figure~\ref{fig:temp}. The lower panel of Figure~\ref{fig:temp} shows the evolution of the SED-fitted temperature of AT~2019abn, assuming that the extinction does not evolve during this part of the event. In Figure~\ref{fig:temp}, we also illustrate the temperature evolution when taking reddening values of either $E_{B-V}=0.79$ or $E_{B-V}=0.90$ (again, see Section~\ref{ext}). Our fitting is done using the specific filter responses and CCD quantum efficiency of the LT (see \citealp{smith_steele_2017}).

After appearing to stay approximately constant while AT~2019abn is around optical peak, the temperature then declines approximately linearly with time until the end of our observations. A linear fit to all of the temperature data at times $t>30$\,days indicates a decline of $25\pm3$\,K\,day$^{-1}$. The uncertainty on this temperature decline rate is dominated by the assumed reddening (i.e.\ between $E_{B-V}=0.79$ and 0.90, which in turn is tied to the assumption made regarding the temperature at peak optical brightness).

\section{Extinction}\label{ext}
Given the implied temperature from the spectra of ILRTs around peak brightness (i.e.\ similar to that of an F-star) combined with the extremely red colour, it is clear that AT~2019abn suffers from significant dust extinction. This can broadly be split into three categories: Galactic-interstellar, M51-interstellar, and circumstellar.

\begin{figure}
\centering
\includegraphics[width=\columnwidth]{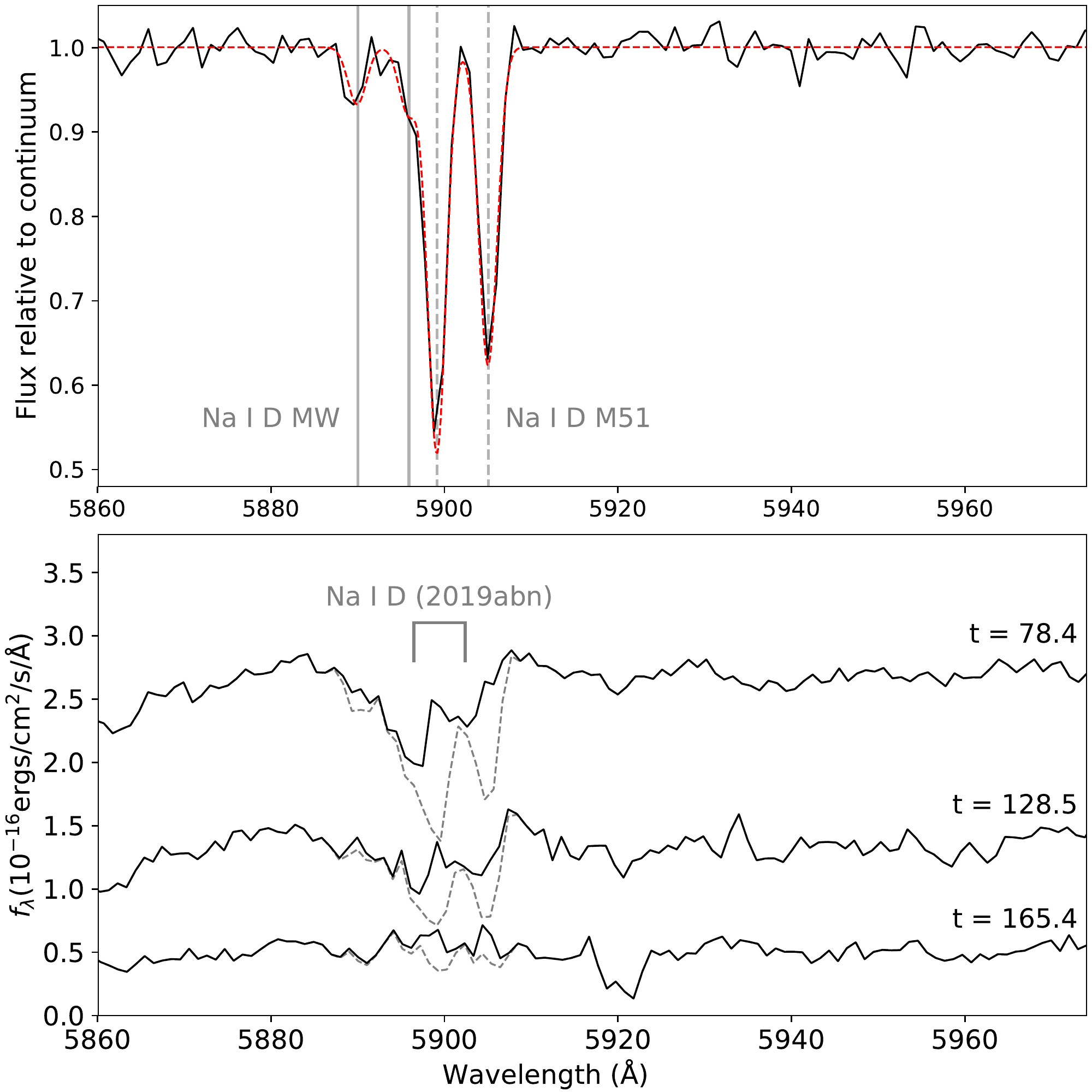}
\caption{\textit{Top:} Region around Na~{\sc i}\,D from the first GTC spectrum, with the Galactic and M51 components indicated. The Gaussian fits of the Galactic and M51 components are shown by the red dashed line. \textit{Bottom:} Solid black lines show the spectra for each of the second, third, and fourth GTC epochs, divided by the fitted Na~{\sc i}\,D profile shown in the upper panel (corrected for the $+0.4$\,\AA\ shift discussed in the text) . This removes the effect from line-of-sight absorption and reveals higher velocity Na~{\sc i}\,D absorption, associated with the outburst of AT~2019abn itself. There is no evidence for such absorption in the final spectrum, however. In each case, the uncorrected spectrum is shown by the dashed grey lines.\label{nai}}
\end{figure}

\citet{2011ApJ...737..103S} find Galactic reddening of $E_{B-V}=0.03$\,mag toward M51. The redshift of M51 allows us to separate the Galactic and M51 Na~{\sc i}\,D interstellar lines in the GTC spectra. From the first GTC spectrum, we measure the Galactic absorption from Na~{\sc i}\,D$_2$ 5890.0\,\AA\ to have an equivalent width (EW) of $0.18\pm0.04$\,\AA, corresponding to $E_{B-V}\sim0.03$\,mag \citep{2012MNRAS.426.1465P}, consistent with that derived from the \citet{2011ApJ...737..103S} dust maps. The Galactic Na~{\sc i}\,D$_1$ 5895.9\,\AA\ line is blended with the much stronger M51 Na~{\sc i}\,D$_2$ absorption. The different components of Na~{\sc i}\,D absorption are illustrated in Figure~\ref{nai}. The Galactic reddening is therefore well constrained and we adopt the value of $E_{B-V}=0.03$ for this work.

We measure EW~=~$1.00\pm0.03$\,\AA\ for the M51 Na~{\sc i}\,D$_1$ absorption. At such high EW values, Na~{\sc i}\,D is saturated and no longer gives a meaningful constraint on the reddening \citep{1997A&A...318..269M}. We can also measure M51 diffuse interstellar bands (DIBs) and find 5778+5780\,\AA\ DIB EW = 0.32\,\AA. This is at the point where many DIB measurements also poorly constrain the reddening, with this 5778+5790 DIB indicating M51 reddening of $E_{B-V}>0.4$\,mag \citep{2015MNRAS.452.3629L}.

In order to constrain the reddening from M51-ISM, plus CSM, we must make some assumptions for temperature. If we follow \citet{2019arXiv190407857J} and assume a peak temperature of 7500\,K, fitting a blackbody to our $BVr'i'z'H$ epoch closest to optical peak, we derive a reddening of $E_{B-V}=0.85\pm0.03$. The reddening will be of course be sensitive to the assumed temperature. If we alternatively assume a temperature of 7000 and 8000\,K, we derive reddening of $E_{B-V}=0.79\pm0.03$ and $E_{B-V}=0.90\pm0.03$ respectively, all assuming $R_V=3.1$. These values are also calculated using the specific filter responses and CCD quantum efficiency of the LT.

\section{Spectroscopic Evolution}
\begin{figure}
\centering
\includegraphics[width=\columnwidth]{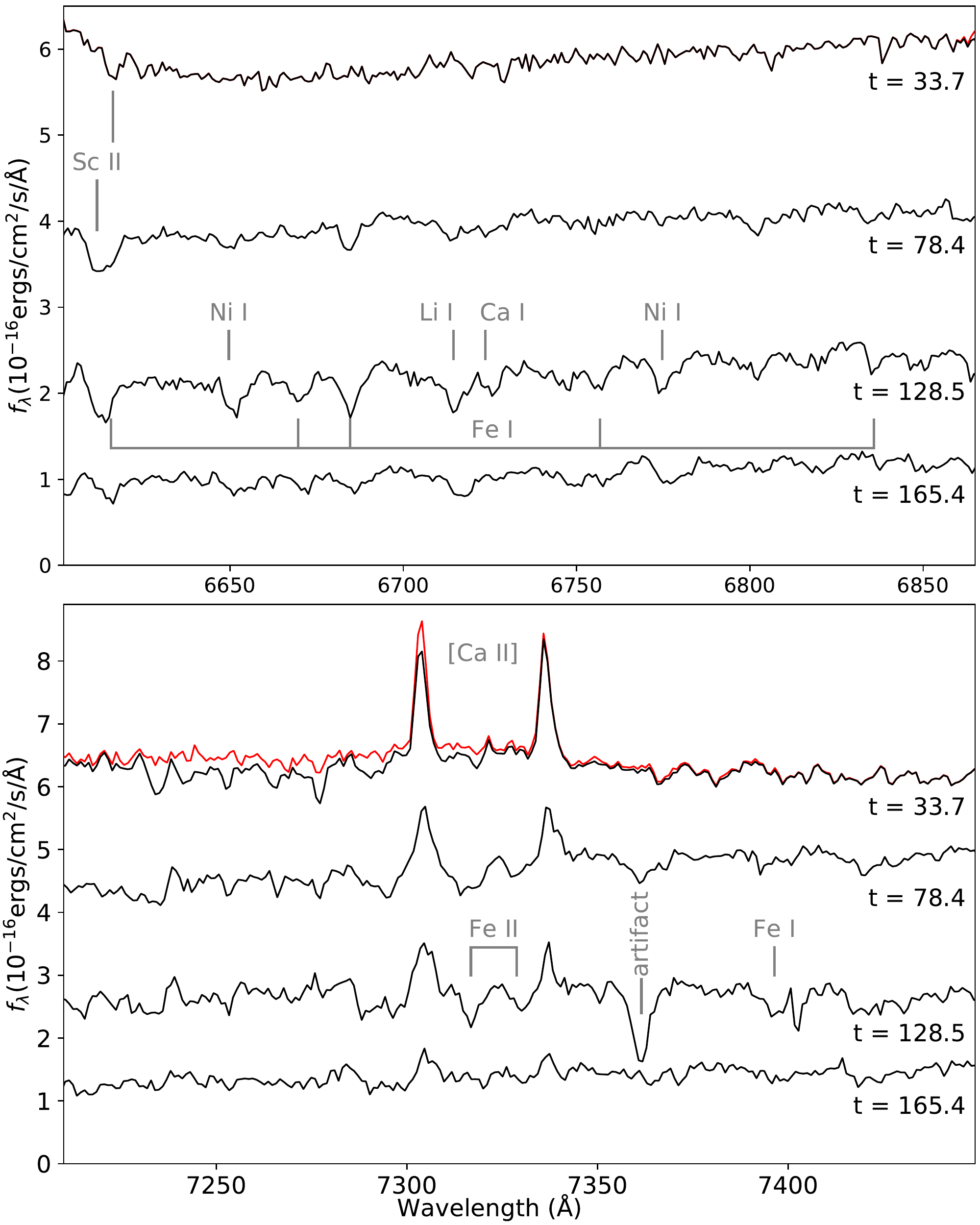}
\caption{Sections of the GTC spectra illustrating the appearance and evolution of absorption lines, along with the evolution of the [Ca~{\sc ii}] emission lines. The calibrated spectra are shown in black, with the telluric-corrected $t=33.7$\,d spectrum shown in red.}
\label{fig:metal}
\end{figure}

The spectra of AT~2019abn are very similar to other ILRTs, with the strongest features being Na~{\sc i}\,D absorption, along with H$\alpha$ and [Ca~{\sc ii}] (7291 and 7324\,\AA) in emission. The spectra evolve to cooler temperatures, with singly ionised and neutral metal absorption lines appearing. This is illustrated in Figure~\ref{fig:metal}. The full series of LT and GTC spectra are shown in Figures~\ref{fig:sprat} and \ref{fig:osiris}.

\subsection{Narrow, low-velocity absorption lines} \label{sec:narrow}
\begin{figure}
\centering
\includegraphics[width=\columnwidth]{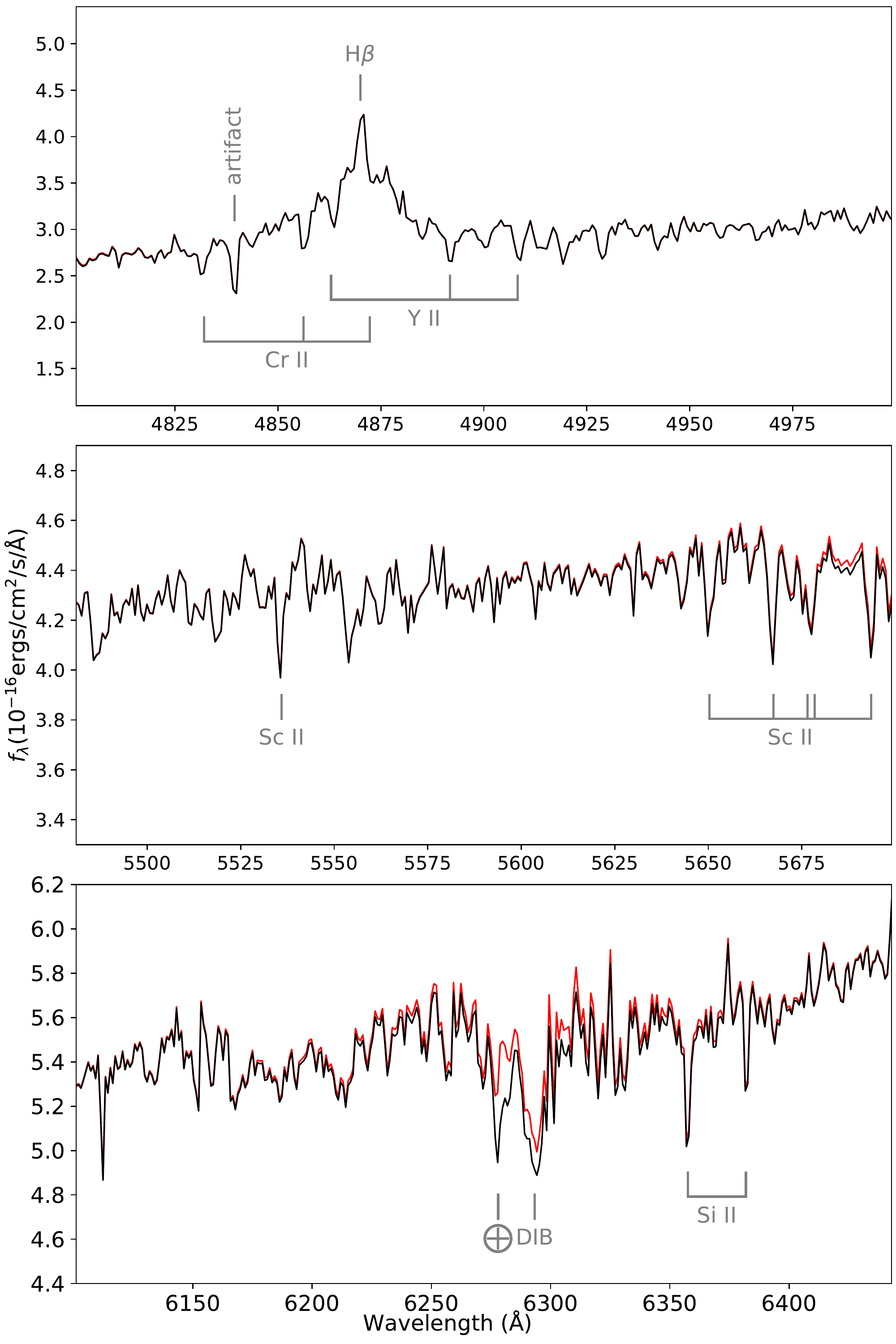}
\caption{Sections of the $t=33.7$\,d GTC spectrum of AT~2019abn, clearly showing narrow, low-velocity absorption from species such as Cr~{\sc ii}, Sc~{\sc ii}, Si~{\sc ii} and Y~{\sc ii}. The calibrated observed spectrum is shown in black, with the telluric-corrected spectrum in red (this only significantly affects the bottom plot).}
\label{fig:low}
\end{figure}

Our first GTC spectrum, taken 33.7\,days after discovery, as the transient had just passed peak optical luminosity, shows low-velocity, narrow absorption lines, which are displayed in Figure~\ref{fig:low}. The lines are unresolved in our spectra, indicating FWHM~$<$~120\,km\,s$^{-1}$. These features are seen clearly for Cr~{\sc ii}, Sc~{\sc ii}, Si~{\sc ii} and Y~{\sc ii}. Na~{\sc i}\,D could well be present, but would be swamped by persistent narrow Na~{\sc i}\,D absorption seen throughout all spectra. These other lines cannot be interstellar in origin (which presumably is the case for at least some fraction of the Na~{\sc i}\,D absorption) as they are not resonance lines. Tracing the velocity evolution of the Sc~{\sc ii} absorption lines (which are visible in most of the spectra) shows the velocity dramatically increasing between $t=33.7$ and $t=78.4$\,days (see top panel of Figure~\ref{fig:metal}). It is hard to reconcile these early low-velocity absorption features with the outflow or ejecta. As in this scenario, the region where the absorption lines are produced would need to move to dramatically higher velocities as the object initially fades from peak brightness.

We instead interpret these absorption lines as arising from a different component. This seems most likely to be pre-existing circumstellar material, ejected from the star either in a stellar wind or prior outburst. The implied velocity of this absorbing material is low. The best-fit radial velocity for the Si~{\sc ii} lines, which have the highest signal-to-noise (S/N) and are least likely to be contaminated, with respect to the M51 Na~{\sc i}\,D absorption is $\sim$\,$+55$\,km\,s$^{-1}$. At the same epoch, the best-fit velocity of the Sc~{\sc ii} lines is of similar order at $\sim$\,$+43$\,km\,s$^{-1}$. The uncertainties on these velocities will be dominated by the uncertainty in the wavelength calibration. While in principle such redshift could imply inflowing gas, it could simply be due to the random motion of the AT~2019abn system with respect to the M51-interstellar dust, which may dominate the Na~{\sc i}\,D absorption at this epoch. It is worth noting that narrow, low-velocity absorption lines were also seen in NGC~300~OT \citep{2009ApJ...699.1850B,2011ApJ...743..118H}.

In addition to the narrow lines, we observe a much broader absorption line at 6294\,\AA\, shown in the bottom panel of Figure~\ref{fig:low}. This is near a region of O$_2$ telluric absorption. As this first spectrum has high S/N and a much clearer continuum than the later GTC spectra (as it is not affected by metal absorption to any large degree), a good telluric correction can be made. We corrected the spectrum for O$_2$ and H$_2$O absorption by fitting for those molecules in the regions $6800-7000$ and $7215-7285$\,\AA\ in \texttt{Molecfit} \citep{2015A&A...576A..78K,2015A&A...576A..77S}. The best-fit O$_2$ and H$_2$O absorption from those regions was then applied to the entire spectrum. This telluric-corrected spectrum is also shown in Figure~\ref{fig:low}. It can be seen from this that telluric absorption can explain the absorption feature blue-ward of 6294\,\AA, but the 6294\,\AA\ absorption line itself remains. There is also tentative evidence for such a feature in the highest S/N LT SPRAT spectra (see Figure~\ref{fig:sprat}). This line is probably due to the 6283\,\AA\ DIB in M51. 

\subsection{Velocity evolution} \label{sec:vel}
Ignoring the potential CSM origin of the early low-velocity lines, we see a move from higher to lower velocities for the absorption lines between the third and fourth GTC spectra. We simultaneously fit the spectral region $6600-6860$ with multiple Gaussian absorption lines, corresponding to Sc~{\sc ii}, Fe~{\sc i}, Ca~{\sc i}, Li~{\sc i,} and Ni~{\sc i,} with the offsets set between the different lines fixed with respect to each other in velocity-space and a single FWHM (in velocity space) for all the lines. This yields best-fitting absorption minimum velocities of $-166$, $-149,$ and $-76$\,km\,s$^{-1}$ (taking $z=0.00154$ as $0$\,km\,s$^{-1}$) for the 78.4, 128.5, and 165.4\,day spectra, respectively. The uncertainties on these velocities will be dominated by the continuum fitting, which is difficult with so many absorption lines present, and the uncertainty on the wavelength calibration of the spectra. We therefore estimate the errors to be $\sim$\,20\,km\,s$^{-1}$.

The Na~{\sc i}\,D complex fit of the first GTC spectrum (shown in Figure~\ref{nai}) initially came out slightly bluer than expected at $z=-0.00006\pm0.00004$ and $z=0.00148\pm0.00001$ for the two components. Correcting the subsequent GTC spectra for this fitted Na~{\sc i}\,D absorption left some residual absorption at the red end of Na~{\sc i}\,D 5895.9\,\AA, visible in all three spectra. If we shift the spectra by $+$\,0.4\,\AA, thereby placing the Galactic Na~{\sc i} lines at $z=0$ and that of M51 to $z=0.00155$, essentially identical to the canonical value of $z=0.00154$, the previously discussed residual then disappears, so we interpret this as a systematic error in the wavelength calibration. The region around Na~{\sc i}\,D, corrected for the narrow Na~{\sc i}\,D complex (Galactic and M51; fit from the first spectrum) is shown in Figure~\ref{nai}. This reveals that after the first spectrum, we also see higher velocity Na~{\sc i}\,D absorption associated with the outflowing or ejected material. This is already visible before the correction as we can see a blue-ward broadening of the lines. This additional Na~{\sc i}\,D absorption is at consistent velocities to other lines, such as Ba~{\sc ii} and Sc~{\sc ii}.

In the second, third, and fourth GTC spectra, the forest of metal absorption lines makes accurately measuring the width of the [Ca~{\sc ii}] lines difficult, including some cases where lines are superimposed, but there is no clear evidence of [Ca~{\sc ii}] velocity evolution during this time. The observation that [Ca~{\sc ii}] does not mirror the absorption line evolution is unsurprising given the lines must be produced in the low density region (to avoid collisional de-excitation).

\subsection{Emission lines}
\begin{figure}
\centering
\includegraphics[width=\columnwidth]{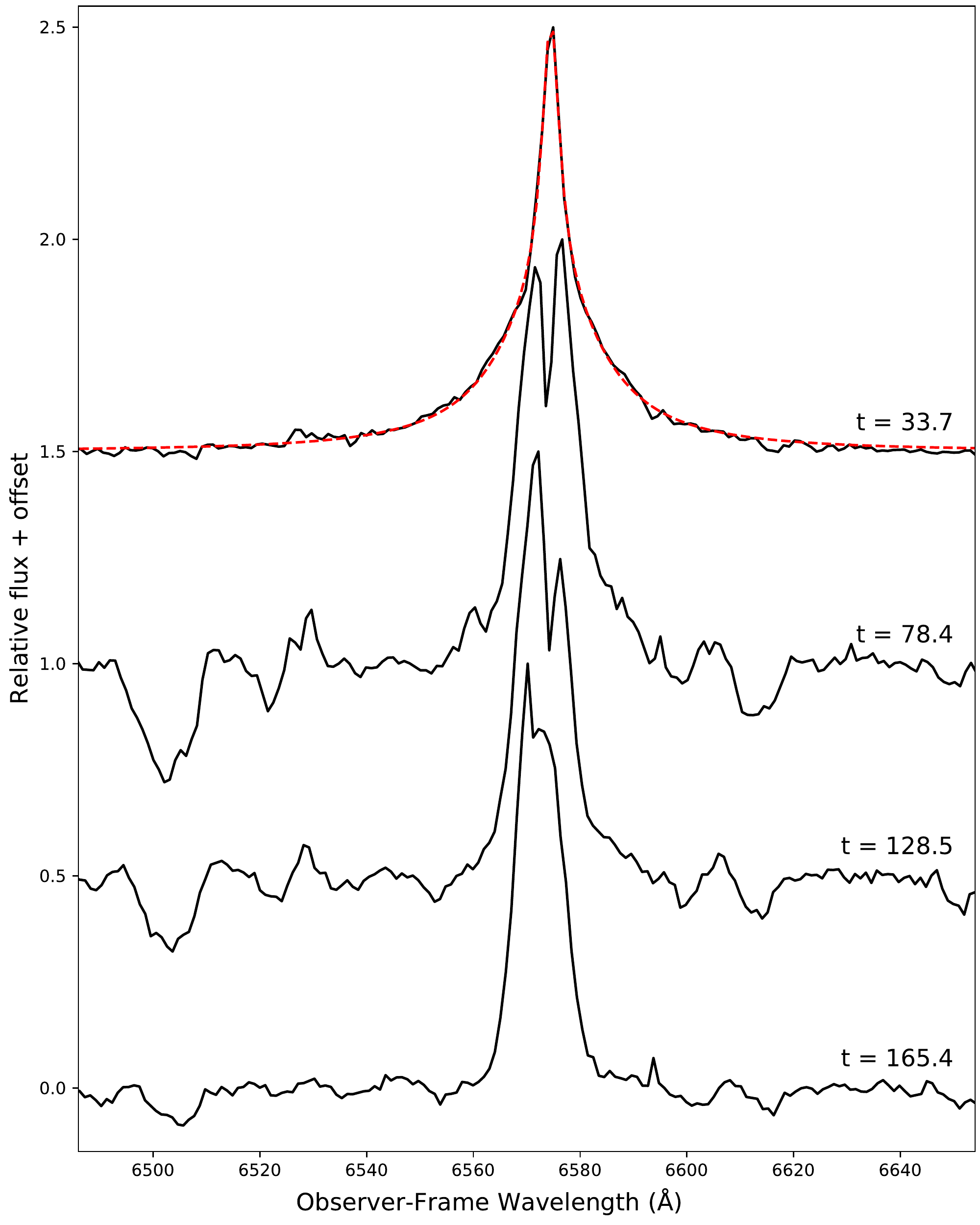}
\caption{Evolution of the H$\alpha$ emission line of AT~2019abn in our four GTC spectra. The red dashed line shows a two-component (Lorentzian + Gaussian) fit to the $t=33.7$\,day spectrum.}
\label{fig:ha-morph}
\end{figure}

In the day~33.7\,d spectrum, the high S/N H$\alpha$ line shows a narrow peak with broad wings. The line is poorly fit by either a single Gaussian or single Lorentzian profile. A two-component Lorentzian profile gives a better fit to the data than a two-component Gaussian profile. However, given the narrow component of the fit is only just resolved, we opt to fit this component with a Gaussian. The resulting broad-Lorentzian + narrow-Gaussian fit is shown in Figure~\ref{fig:ha-morph}. After correcting for spectral resolution, we measure a FWHM of $\sim$\,900\,km\,s$^{-1}$ for the broad component and a FWHM of $\sim$\,130\,km\,s$^{-1}$ for the narrow component. The narrower component is approaching the spectral resolution ($R\sim2500$), which should be kept in mind when considering this $\sim$\,130\,km\,s$^{-1}$ velocity measurement.

Fitting the H$\alpha$ and H$\beta$ lines in the first GTC spectrum indicates a redshift of $z\sim0.00178$. This is higher than the typical redshift value for M51 and could potentially be influenced by asymmetric lines. High-resolution spectroscopy of NGC~300~OT showed some of the emission lines to be highly asymmetric, with the blue side of the [Ca~{\sc ii}] lines almost entirely suppressed \citep{2009ApJ...699.1850B}. At a lower resolution, this could still give a reasonable line fit, but with the effect of having the lines appearing to be slightly redshifted. It is worth noting however, that this higher redshift of $z\sim0.00178$ (i.e.\ $\sim$\,70\,km\,s$^{-1}$ from the redshift of M51), is similar to the observed `redshift' of the early narrow absorption lines discussed in Section~\ref{sec:narrow}, possibly pointing towards a peculiar velocity of AT~2019abn with respect to the dust causing the majority of the Na~{\sc i}\,D absorption. If this is the case, then the absorption-line velocities discussed in Section~\ref{sec:vel} would be $\sim$\,60\,km\,s$^{-1}$ higher (i.e.\ more negative).

In the latter three GTC spectra, the H$\alpha$ line shows more structure. Some of this double-peaked structure is real and has been seen in other ILRTs (e.g.\ NGC~300~OT; \citealp{2009ApJ...699.1850B}). However, it is not possible to quantitatively fit these components due to H$\alpha$ emission in the surrounding region of M51, which makes an accurate background subtraction difficult. Given all of our spectra were taken at the parallactic angle, it is also possible that the amount of background contamination varies between spectra. The double peaked structure of the H$\alpha$ line indicates asphericity and could even point to a disk-like configuration (see e.g.\ \citealp{2000ApJ...536..239L,2018MNRAS.477...74A}). The fact that this is seen only at late times could point towards it being embedded in a more spherical ejecta initially and then becoming revealed when the optical depth of the more spherical component becomes sufficiently low.

\subsection{Temperature from spectra}
The absorption line evolution from the GTC spectra is broadly consistent with that implied from the post-peak SED fitting, where the material gradually cools. Assuming the peak-luminosity low-velocity absorption lines are associated with pre-existing material as discussed in Section~\ref{sec:narrow}, at peak, the photosphere is too hot to produce strong metal absorption lines. Around 45 days later, when the second GTC spectrum is taken, the ejecta have cooled sufficiently for a host of singularly ionised metal lines to appear, along with some neutral metal lines. The photometric SED fitting implies that during this time, the photosphere could have cooled by $\sim$1000\,K.

When the next GTC spectrum was taken, another 50\,days later, the SED fitting indicates that the material has cooled by a further $>$1000\,K, which is reflected by more prominent neutral metal absorption, particularly Fe~{\sc i} and Ni~{\sc i}. Our SED fitting assumes no dust formation during the phases observed in this work, which, if present, would have the effect of the fitting giving a cooler temperature than was the case. While we cannot accurately measure the temperature from the spectra, they do at least confirm that the observed photosphere is indeed cooling with time.

\section{Discussion}
After correcting for the implied reddening, AT~2019abn is shown to be the most luminous ILRT observed to date (see Section~\ref{sec:com}). At $M_{r'}=-15.2\pm0.2$, it is in the absolute magnitude range of low-luminosity Type~IIP SNe. However, these low-luminosity Type IIP~SNe still have velocities $>$\,1000\,km\,s$^{-1}$ (see e.g. \citealp{2004MNRAS.347...74P,2018ApJ...859...78N}), much higher than anything we observe from AT~2019abn. While AT~2019abn and other ILRTs show similar spectra to LBV outbursts such as UGC~2773~OT~2009-1 and SN~2009ip \citep{2010AJ....139.1451S,2011ApJ...732...32F}, their light-curve evolution is much more rapid that those LBV eruptions.

\citet{2019arXiv190407857J} identified a variable 4.5\,$\mu$m source in archival \textit{Spitzer} data, coincident with the position of AT~2019abn. Dusty progenitor stars were also found for SN~2008S and NGC~300~OT \citep{2008ApJ...681L...9P,2009ApJ...699.1850B,2009ApJ...695L.154B}. However, the data published by \citet{2019arXiv190407857J} for AT~2019abn demonstrate the first time that variability has been detected in the progenitor luminosity. Infrared follow-up of AT~2019abn over the coming decade will be important in helping to understand its nature. Both SN~2008S and NGC~300~OT are now fainter than their progenitor stars \citep{2016MNRAS.460.1645A}.

\subsection{Comparison to other ILRTs} \label{sec:com}

\begin{figure}
\centering
\includegraphics[width=\columnwidth]{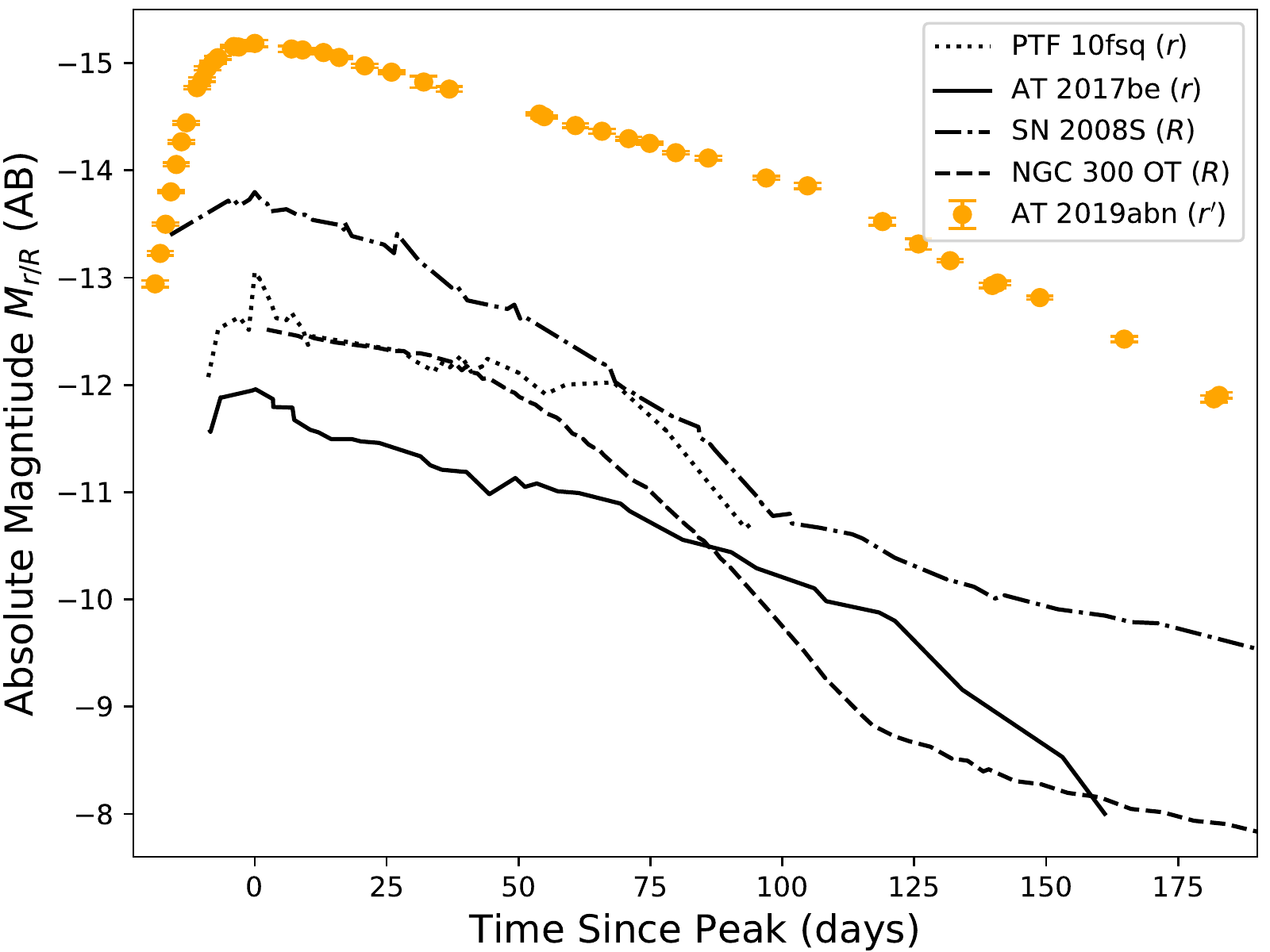}
\caption{Comparison between $r'/r/R$-band light curves of AT~2019abn to SN~2008S, NGC~300~OT, PTF~10fqs and AT~2017be. All light curves have been corrected for Galactic and internal reddening (as discussed in the text).\label{fig:comp}}
\end{figure}

We compare the absolute $r'$-band light curve of AT~2019abn with other ILRTs in Figure~\ref{fig:comp}. The light curves are SN~2008S (assuming $E_{B-V,~ {\mathrm{Host+CSM}}}=0.3$, \citealp{2009MNRAS.398.1041B}; $\mu_0=28.78$, \citealp{2006MNRAS.372.1315S}), NGC~300~OT ($E_{B-V,~ {\mathrm{Host+CSM}}}=0.25$, \citealp{2018MNRAS.480.3424C}; $\mu_0=26.29$, \citealp{2016AJ....151...88B}), PTF~10fqs ($E_{B-V,~ {\mathrm{Host+CSM}}}=0.4$, this work; $\mu_0=30.82$, \citealp{2009ApJ...694.1067P}), and AT~2017be ($E_{B-V,~ {\mathrm{Host+CSM}}}=0.04$, $\mu_0=29.47$; \citealp{2018MNRAS.480.3424C}). All light curves are also corrected for foreground Galactic reddening from the \citet{2011ApJ...737..103S} dust maps.
Figure~\ref{fig:comp} shows that AT~2019abn is substantially more luminous than other members of the class. 

The ($r-i$) colour of PTF~10fqs near peak (see \citealp{2011ApJ...730..134K}) suggests significant reddening. The assumption of significant dust is consistent with the high Na~{\sc i}\,D EW and low SED-fitted temperature of $\sim3900$\,K (when no extinction correction is made; \citealp{2011ApJ...730..134K}). Given the spectroscopic similarity between PTF~10fqs and other ILRTs, such a low peak temperature seems implausible (indeed \citealp{2011ApJ...730..134K} only derive this temperature as a lower limit). Assuming a similar temperature to other ILRTs and $R_V=3.1$, the $(r-i)$ colour suggests additional reddening in the region of $E_{B-V}\sim0.4$\,mag for PTF~10fqs. We use this value to correct the $r$-band light curve of PTF~10fqs, as discussed above.

Despite AT~2017be and AT~2019abn being the lowest and highest luminosity ILRTs respectively, the shape of their \textit{r}-band light curves are very similar, with both displaying a fast rise to peak, a linear (in magnitude) slow decline, which was then followed by a faster linear decline. The exception to this was AT~2017be showing a faster decline shortly after peak, prior to the slow linear decline described above \citep{2018MNRAS.480.3424C}.

\subsection{H$\alpha$ flux evolution}
\begin{figure}
\centering
\includegraphics[width=\columnwidth]{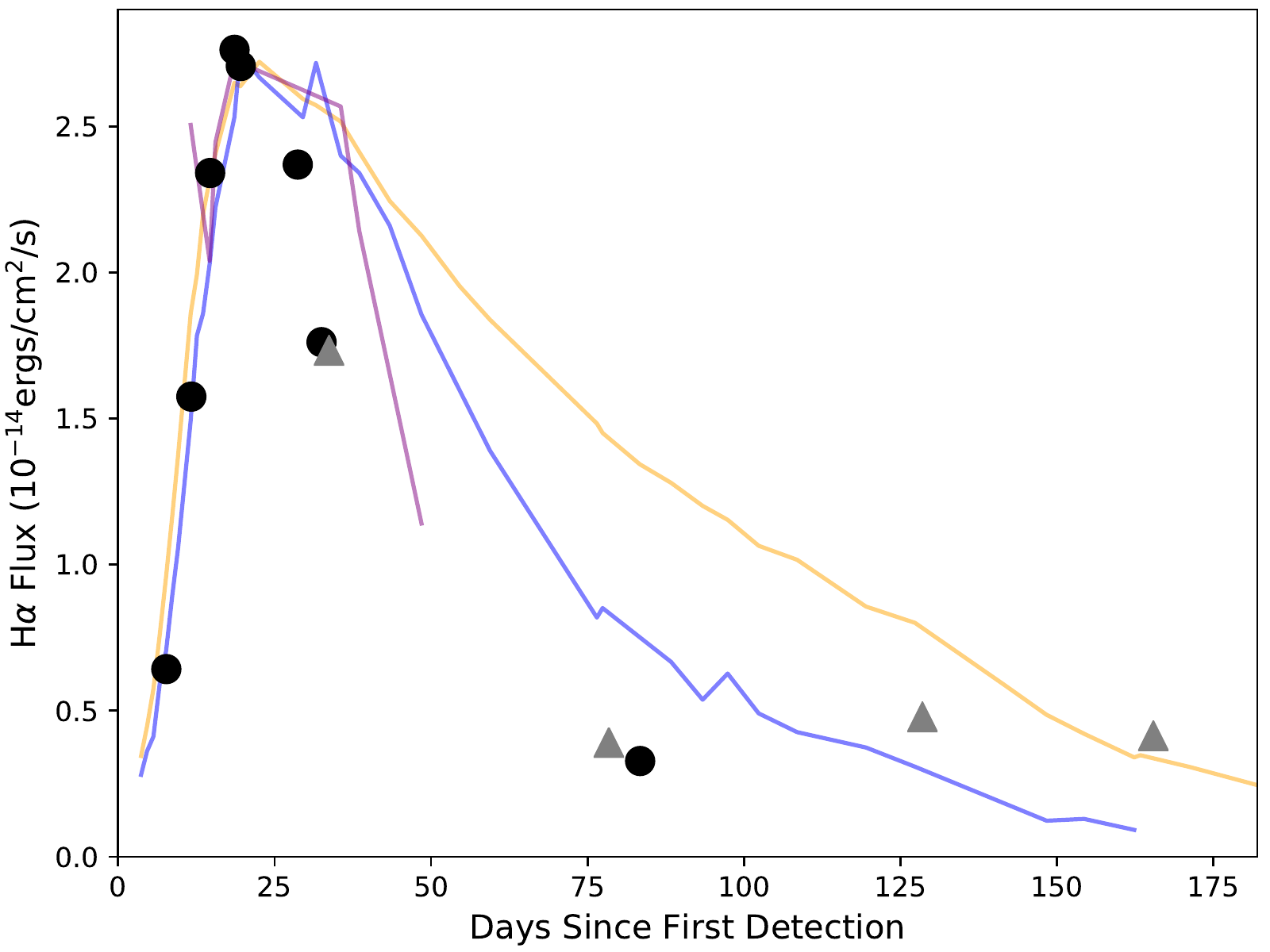}
\caption{Evolution of H$\alpha$ emission flux for AT~2019abn. Black circles show SPRAT measurements, with grey triangles showing OSIRIS measurements. The uncertainties on the calibration and measurements themselves will generally be relatively low. In some cases there may be significant uncertainties from the sky subtraction however, due to the complicated M51 background. The \textit{u}$'$-band (purple), \textit{B}-band (blue) and \textit{r}-band (orange) optical light curves are also shown for reference (in relative flux, rather than magnitude, and not to any absolute scale).}
\label{fig:ha}
\end{figure}

The evolution of the H$\alpha$ emission-line flux for AT~2019abn is shown in Figure~\ref{fig:ha}. This shows that H$\alpha$ emission peaks at around the same time as the optical continuum. However, the H$\alpha$ flux declines much more rapidly than the optical continuum, even when compared to the \textit{B}-band. The \textit{u}$'$-band decline could be similar but the combination of poor \textit{u}$'$-band coverage (due to the faintness of AT~2019abn in that band) and lack of spectra between day~34 and day~79 makes it impossible to tell. From our spectra taken from $t=78.4$ onward, the H$\alpha$ flux appears to remain approximately constant. Similar behaviour of a rapid H$\alpha$ decline followed by a plateau has been seen in other ILRTs (see Fig.~11 of \citealp{2018MNRAS.480.3424C}).

\subsection{Temperature and luminosity evolution}
\begin{figure}
\centering
\includegraphics[width=\columnwidth]{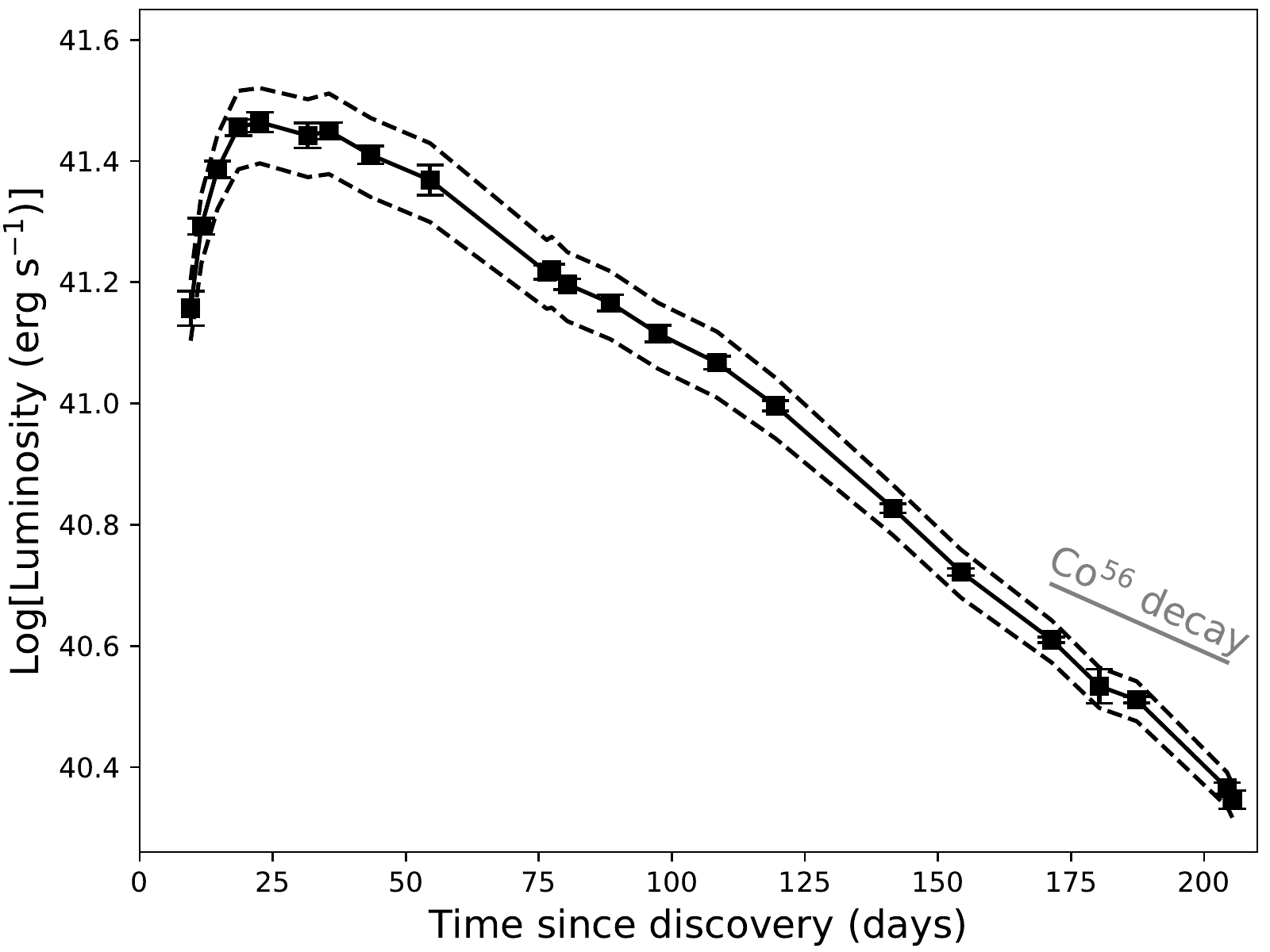}
\includegraphics[width=\columnwidth]{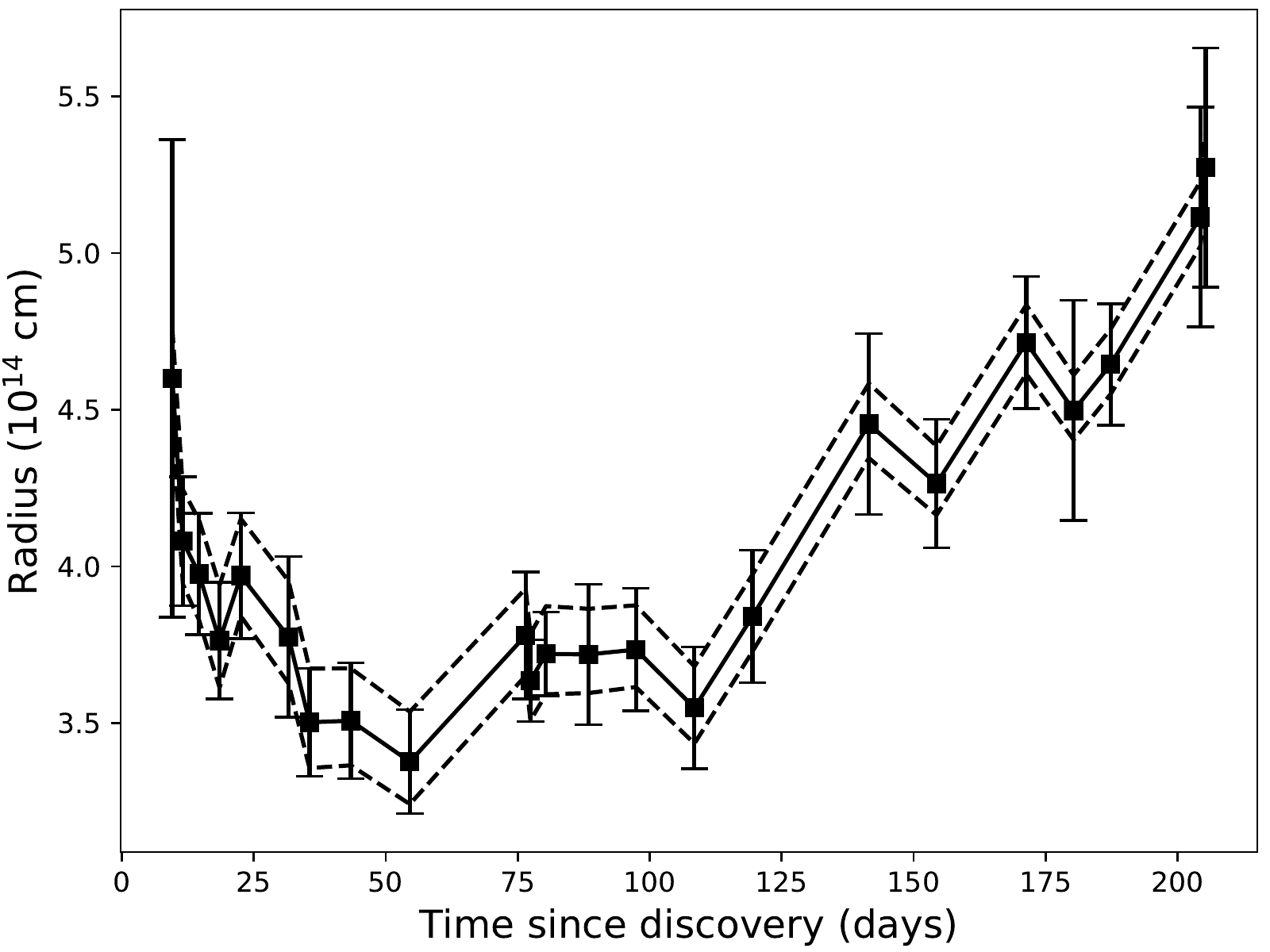}
\caption{\textit{Top:} Luminosity evolution of AT~2019abn, derived from the blackbody fits to the photometry and integrated between 3000--23000\,\AA. The fading timescale expected if the light curve was powered by the radioactive decay of $^{56}$Co is also indicated for reference. \textit{Bottom:} Radius evolution of AT~2019abn, as derived from the luminosity and temperature fits, assuming luminosity, $L=4\pi R^2\sigma{T_{\mathrm{eff}}}^{4}$. The data points, which are joined by solid lines represent $E_{B-V}=0.85$, with the dashed lines representing $E_{B-V}=0.9$ and~0.79.\label{fig:lum}}
\end{figure}

Using our photometry and the fitted temperatures from Figure~\ref{fig:temp}, we compute the integrated luminosity between 3000--23000\,\AA\ (i.e.\ approximately \textit{u}--$K_{\mathrm{s}}$-band), assuming a blackbody. The calculated luminosity evolution is shown in Fig.~\ref{fig:lum}. After turning over at peak, the luminosity evolution follows a monotonic decline, as seen in other ILRTs \citep{2018MNRAS.480.3424C}. There is no evidence for a secondary peak as seen in some LRNe. The rate of decline at the end of our light curve rules out radioactive decay of $^{56}$Co as the primary energy source powering the light curve up to the end of our observations. This monotonically declining luminosity is similar to that seen in other ILRTs, and in sharp contrast to LRNe, which show a long plateau or second peak in their luminosity. This plateau or rise to secondary peak in LRNe has already started by 40\,days after peak brightness (see Fig.~4 of \citealp{2019arXiv190913147C}), yet our data of AT~2019abn over the course of 200\,days ($>$170\,days after peak) show no sign of such a signature.

We also compute the radius evolution from the luminosity and temperature calculations, assuming luminosity, $L=4\pi R^2\sigma{T_{\mathrm{eff}}}^{4}$, where \textit{R} is the radius and $\sigma$ is the Stefan–Boltzmann constant. This evolution is shown in the bottom panel of Figure~\ref{fig:lum}. The radius initially seems to fall before gradually increasing. This appears different from the typical ILRT behaviour, which shows a slowly declining radius; whereas LRNe show radius increasing with time (see Fig.~4 of \citealp{2019arXiv190913147C}). However, if we look in more detail, this also differs from LRN behaviour which show an increase in radius of as much as an order of magnitude by $t=200$\,days (again, see Fig.~4 of \citealp{2019arXiv190913147C}), whereas the difference between the maximum and minimum computed radius for the entire span of observations is only a factor of $\sim1.5$. At the end of our observations (when AT~2019abn became unobservable due to the Sun), AT~2019abn had a magnitude of $V\sim21.5$ and $r'\sim20$, so deep, late-time observations will be needed to track the radius evolution further. The increase in radius from peak optical brightness translates to an increase of $\sim$100\,km\,s$^{-1}$, which is of the same order as the absorption line velocities we observe in the spectra.

\subsection{Rise to peak}
\begin{figure}
\centering
\includegraphics[width=\columnwidth]{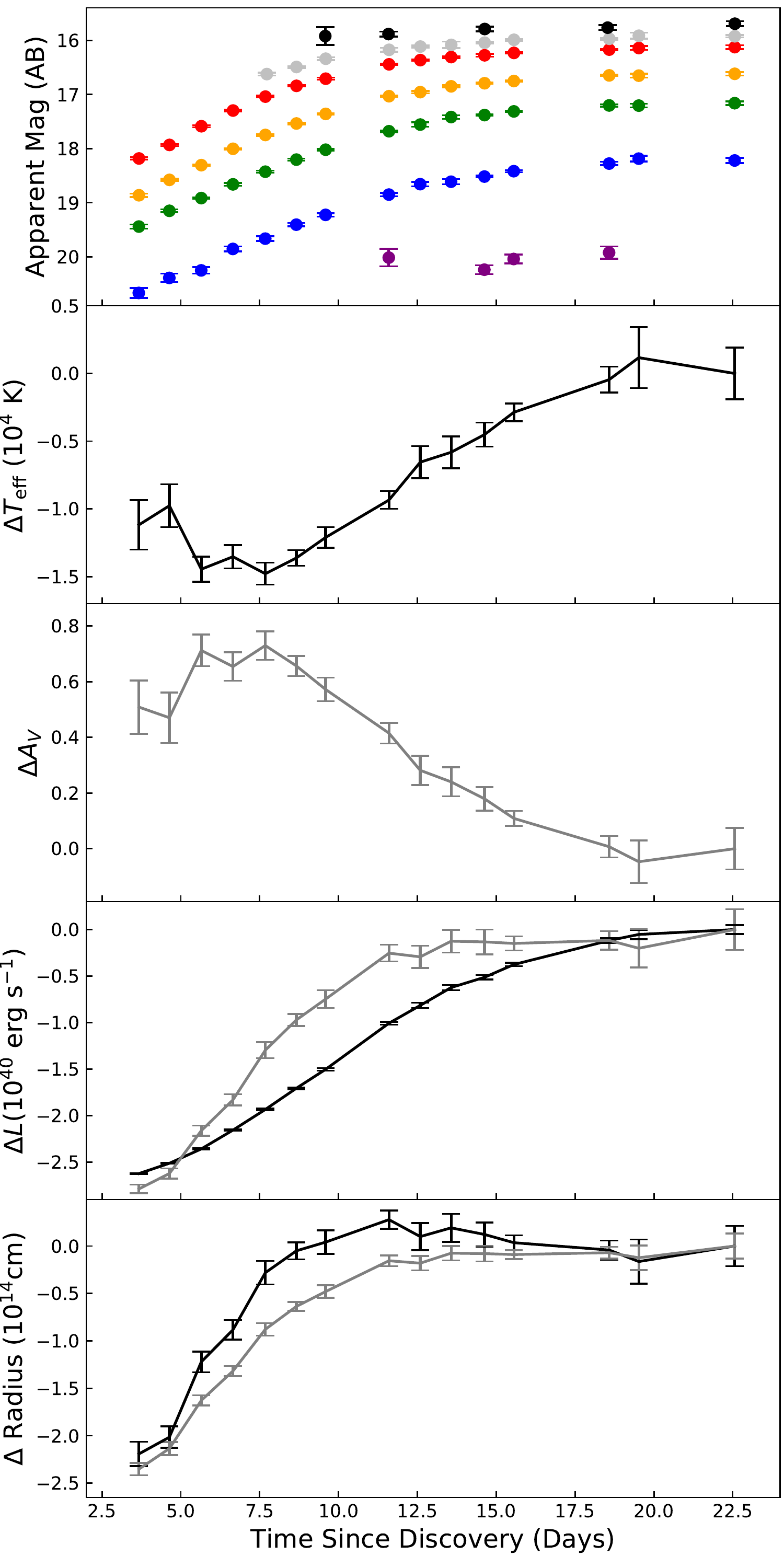}
\caption{\textit{First panel (top):} Early light curve of AT~2019abn (key as in Figure~\ref{fig:lc}). \textit{Second panel:} Early temperature evolution of AT~2019abn, assuming no change in dust extinction. \textit{Third panel:} Implied additional dust extinction to that seen at peak, if one assumes a constant temperature of $T=7500$\,K over this portion of the light curve. \textit{Fourth panel:} early luminosity evolution of AT~2019abn. \textit{Fifth panel (bottom):} early radius evolution of AT~2019abn. In the last two panels, the black and grey points represent the different assumptions of constant extinction ($E_{B-V}=0.85$) and constant temperature ($T=7500$\,K), respectively. \label{early}}
\end{figure}

The early discovery of AT~2019abn and our daily cadence multi-colour follow-up at early times makes it possible to probe the early evolution of an ILRT for the first time. We therefore show the early portion of the light curve in more detail in Figure~\ref{early}. We also fit the the $BVr'i'$ photometry with a blackbody during this stage of the evolution to derive changes in the temperature, dust, luminosity, and radius that are implied by different assumptions. These fits are also shown in Figure~\ref{early}.

Assuming a constant temperature of 7500\,K, \citet{2019arXiv190407857J} suggest dust destruction as the cause of AT~2019abn becoming increasingly blue during its rise to peak. Using the same assumption of constant temperature on our data shows that if all of the pre-peak colour change is due to changes in extinction, additional extinction of $A_V\sim0.7$\,mag (assuming constant $R_V=3.1$) is required to explain the observed colours in the first $\sim$10\,days after discovery. This is consistent with the result derived by \citet{2019arXiv190407857J}. The evolution of this as AT~2019abn rises to peak can be seen in the middle panel of Figure~\ref{early}.

Unfortunately, our early-time data are not sufficient to discriminate between changes in temperature and changes in dust extinction. As an alternative, we therefore assume a fixed extinction (i.e.\ no dust destruction) and we assume that changes in the pre-maximum colour are entirely due to changes in the intrinsic temperature of the photosphere. This is shown in the second panel of Figure~\ref{early}. The implied change in temperature is $\sim$1500\,K between our early data and peak. If similar colour evolution is seen in other ILRTs, high S/N spectra of an ILRT during this early rise should, therefore, help reveal the nature of the behaviour and distinguish between temperature and dust evolution. The bottom two panels of Figure~\ref{early} show the relative early luminosity and radius evolution derived under the two alternative assumptions of constant extinction and constant temperature. Evidence of the approximately achromatic early rise can be seen under both assumptions, where significant changes in the second and third panels of Figure~\ref{early} only really begin around ten\,days after discovery.

\section{Summary and conclusions}
 We conducted multi-wavelength follow-up observations of AT~2019abn, located in the nearby M51 galaxy. Here we summarise our findings:
\begin{enumerate}
\item Our observations of AT~2019abn yield the most detailed early light curve of any ILRT to date, starting around three weeks before and more than two magnitudes below peak brightness.
\item The observations of the initial rise, when AT~2019abn was $>$\,1~mag below peak, are consistent with an achromatic (\textit{BVr$'$i$'$}) rise in luminosity. As it approaches peak, the colours become bluer.
\item AT~2019abn is subject to significant M51 extinction (interstellar + CSM). From the expected peak temperature of an ILRT, we derive an estimate of $E_{B-V}\sim0.85$, assuming $R_V=3.1$.
\item Low-resolution spectroscopy of AT~2019abn, beginning during the rise to peak, shows that the H$\alpha$ flux peaks at a similar time to the optical continuum but then fades much more rapidly than the optical continuum before plateauing.
\item Fitting a blackbody to our multi-wavelength photometry indicates that after peak brightness, the temperature declines slowly over time. We estimate the rate of this decline to be $\sim25$\,K\,day$^{-1}$ during this phase.
\item From the blackbody fits, we find that the luminosity of the transient shows a monotonic decline after peak, similar to other ILRTs and in contrast to LRNe. The fits indicate that the implied radius slowly increases, with this marginal increase over time not matching well with other transients of either the ILRT or LRN class.
\item The GTC spectra taken as AT~2019abn declines are broadly consistent with this temperature evolution, showing increasingly strong absorption from neutral species such as  Fe~{\sc i}, Ni~{\sc i,} and Li~{\sc i}.
\item The first GTC spectrum, taken 33.7\,d after discovery, shows narrow (unresolved) low-velocity absorption from species such as Si~{\sc ii}, Sc~{\sc ii}, Y~{\sc ii,} and Cr~{\sc ii}, which we interpret as most likely arising from pre-existing material from around the system.
\item We conclude that while there may be some differences with other members of the class (such as the radius evolution), AT~2019abn is best described as an ILRT. Observations of the final stages of the transient's evolution will be needed to confirm this evaluation. \end{enumerate}

The early discovery of nearby ILRTs, such as AT~2019abn, will be key in furthering our understanding of these objects, as will building a larger sample with the increased volume in which such objects can be regularly discovered thanks to deeper all-sky surveys such as ZTF, ATLAS, and, in the future, LSST.

\begin{acknowledgements}
We thank the anonymous referee for useful feedback on the submitted manuscript. SCW and IMH acknowledge support from UK Science and Technology Facilities Council (STFC) consolidated grant ST/R000514/1. DJ acknowledges support from the State Research Agency (AEI) of the Spanish Ministry of Science, Innovation and Universities (MCIU) and the European Regional Development Fund (FEDER) under grant AYA2017-83383-P.  DJ also acknowledges support under grant P/308614 financed by funds transferred from the Spanish Ministry of Science, Innovation and Universities, charged to the General State Budgets and with funds transferred from the General Budgets of the Autonomous Community of the Canary Islands by the Ministry of Economy, Industry, Trade and Knowledge. This research was also supported by the Erasmus+  programme  of the European Union under grant number 2017-1-CZ01-KA203-035562. MJD acknowledges support from STFC consolidated grant ST/R000484/1. The work of OP has been supported by Horizon 2020 ERC Starting Grant ``Cat-In-hAT'' (grant agreement \#803158) and INTER-EXCELLENCE grant LTAUSA18093 from the Czech Ministry of Education, Youth, and Sports.

The work is based on observations with the Liverpool Telescope, which is operated on the island of La Palma by Liverpool John Moores University in the Spanish Observatorio del Roque de los Muchachos of the Instituto de Astrofisica de Canarias with financial support from the STFC. Some of these observations were obtained through director’s discretionary time (DDT) programme JQ19A01 (PI: Darnley).

This work is also based on observations made with the Gran Telescopio Canarias (GTC), installed in the Spanish Observatorio del Roque de los Muchachos of the Instituto de Astrofísica de Canarias, in the island of La Palma, under DDT (programme ID GTC2019-110, PI: Jones). 

Also based on observations made with the IAC80 telescope operated on the island of Tenerife by the Instituto de Astrof\'isica de Canarias in the Spanish Observatorio del Teide, and on observations made with the Nordic Optical Telescope, operated by the Nordic Optical Telescope Scientific Association at the Observatorio del Roque de los Muchachos, La Palma, Spain, of the Instituto de Astrof\'isica de Canarias.

\end{acknowledgements}

\bibliographystyle{aa}

\begin{appendix}
\section{LT SPRAT and GTC OSIRIS spectra}
\begin{figure*}
\centering
\includegraphics[width=2\columnwidth]{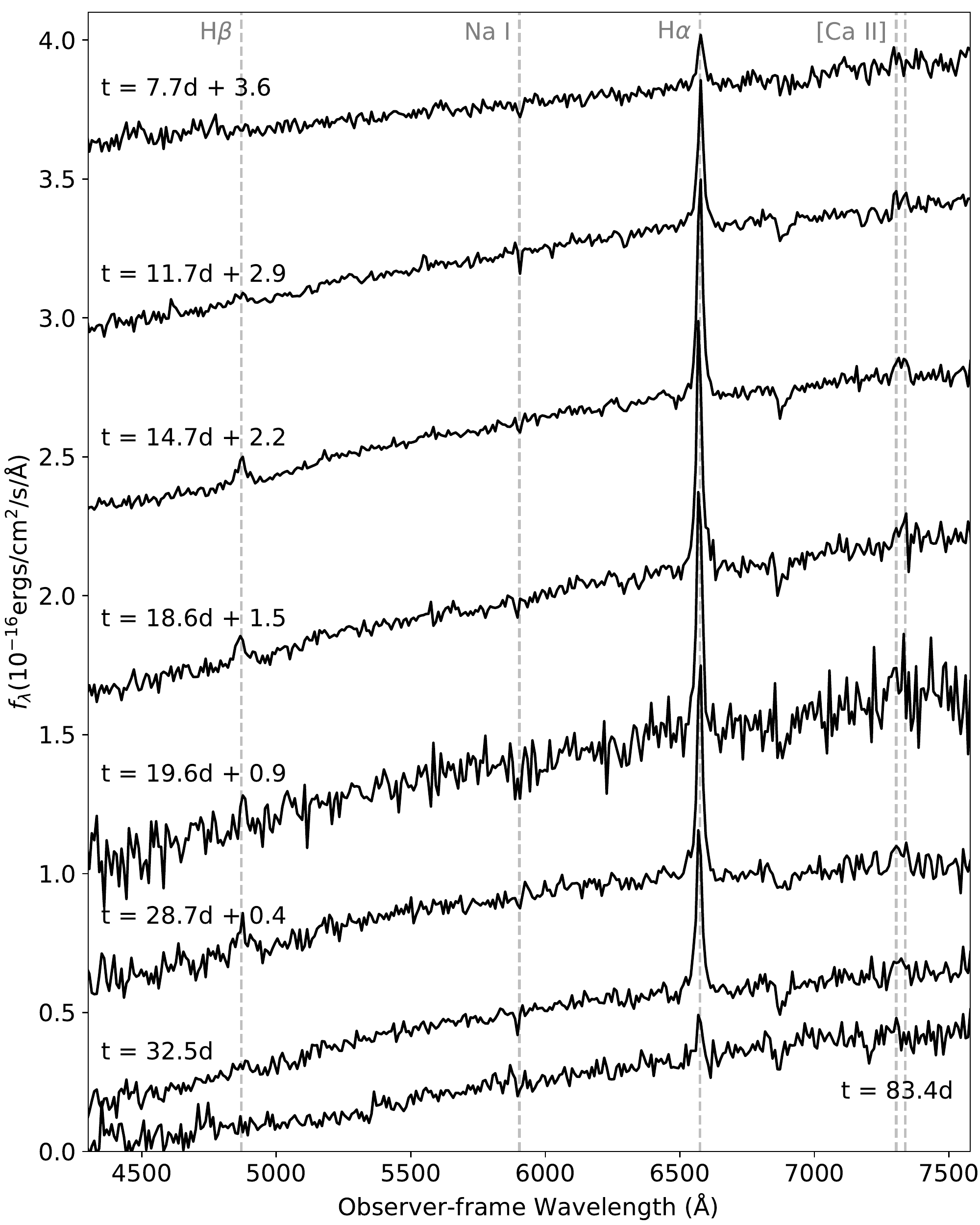}
\caption{All LT Sprat spectra taken of AT~2019abn. These are flux calibrated and include a shift (indicated in the label for each spectrum) for clarity.}
\label{fig:sprat}
\end{figure*}

\begin{figure*}
\centering
\includegraphics[width=2\columnwidth]{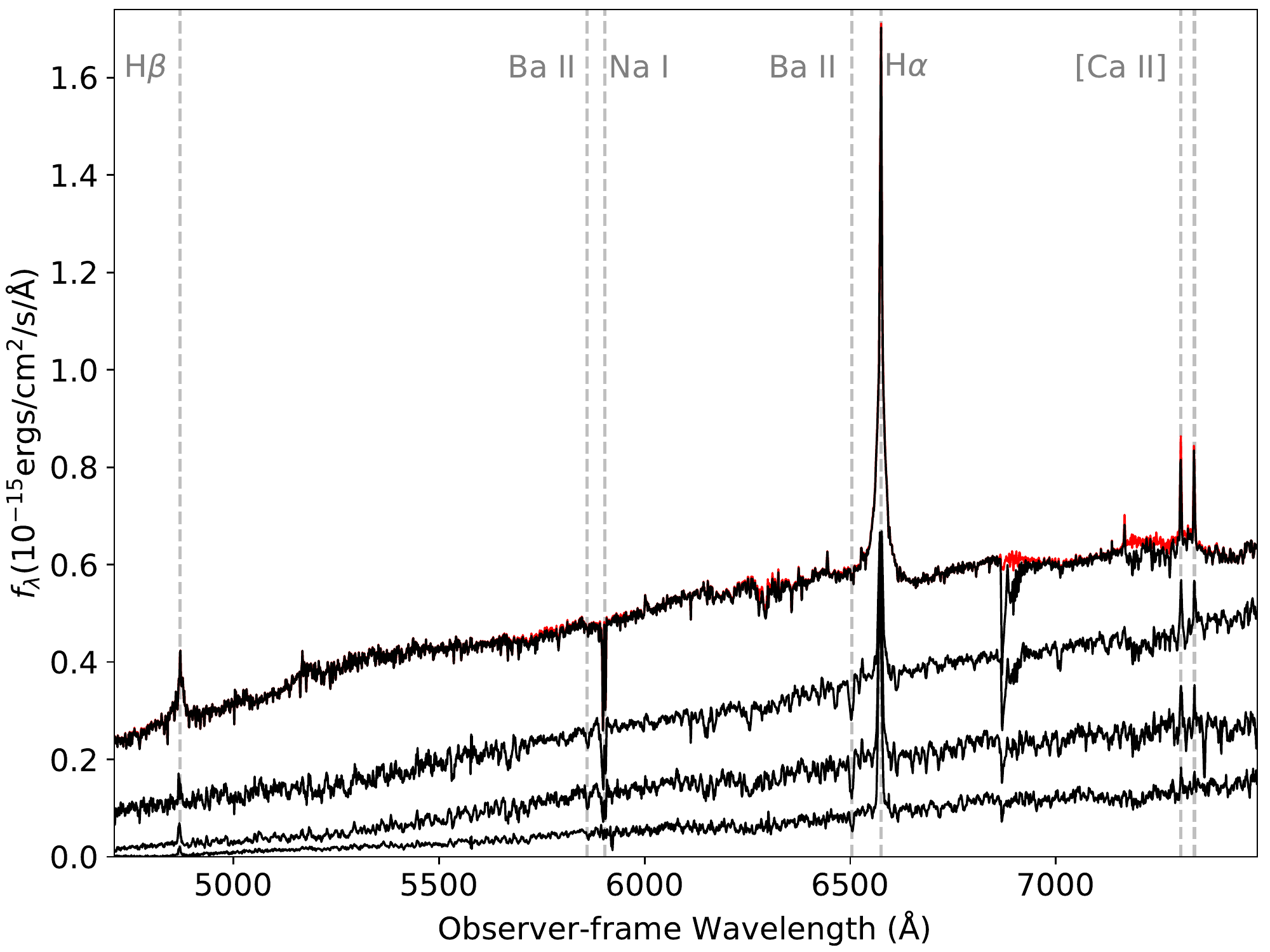}
\caption{GTC OSIRIS spectra of AT~2019abn. Telluric-corrected first spectrum is shown in red. From top to bottom, the spectra were taken at $t=33.7$, 78.4, 128.5, and 165.4\,days, respectively.}
\label{fig:osiris}
\end{figure*}

\clearpage
\section{Photometry (online materials)}.
\centering
\tablefirsthead{\hline\hline MJD &$t$ [days] &Instrument &Filter &Magnitude\\\hline}
\tablehead{\hline\hline MJD &$t$ [days] &Instrument &Filter &Magnitude\\\hline}
\tablecaption{Optical and NIR photometry of AT~2019abn. All measurements are in the AB system and not corrected for any extinction. As in the main text, `$t$' refers to the time since the first optical detection.\vspace{-0.5cm}\label{tab:phot}}
\begin{supertabular}{lcccc}
58517.15 & 11.59 & LT/IO:O & $u'$ & $20.16\pm0.16$\\
58520.18 & 14.62 & LT/IO:O & $u'$ & $20.38\pm0.08$\\ 
58521.11 & 15.55 & LT/IO:O & $u'$ & $20.18\pm0.08$\\ 
58524.13 & 18.58 & LT/IO:O & $u'$ & $20.07\pm0.11$\\ 
58541.13 & 35.57 & LT/IO:O & $u'$ & $20.13\pm0.09$\\ 
58544.09 & 38.53 & LT/IO:O & $u'$ & $20.33\pm0.10$\\ 
58554.03 & 48.47 & LT/IO:O & $u'$ & $21.01\pm0.14$\\ 
\hline
58509.22 & 3.67 & LT/IO:O & $B$ & $20.79\pm0.09$\\ 
58510.19 & 4.63 & LT/IO:O & $B$ & $20.51\pm0.08$\\ 
58511.20 & 5.65 & LT/IO:O & $B$ & $20.37\pm0.06$\\ 
58512.20 & 6.65 & LT/IO:O & $B$ & $19.98\pm0.04$\\ 
58513.23 & 7.68 & LT/IO:O & $B$ & $19.79\pm0.04$\\ 
58514.22 & 8.66 & LT/IO:O & $B$ & $19.53\pm0.03$\\ 
58515.14 & 9.59 & LT/IO:O & $B$ & $19.35\pm0.03$\\ 
58517.15 & 11.60 & LT/IO:O & $B$ & $18.98\pm0.03$\\ 
58518.14 & 12.59 & LT/IO:O & $B$ & $18.78\pm0.04$\\ 
58519.12 & 13.57 & LT/IO:O & $B$ & $18.74\pm0.05$\\ 
58520.18 & 14.63 & LT/IO:O & $B$ & $18.64\pm0.02$\\ 
58521.11 & 15.56 & LT/IO:O & $B$ & $18.54\pm0.02$\\ 
58524.14 & 18.58 & LT/IO:O & $B$ & $18.40\pm0.03$\\ 
58525.07 & 19.52 & LT/IO:O & $B$ & $18.31\pm0.06$\\ 
58528.11 & 22.56 & LT/IO:O & $B$ & $18.34\pm0.05$\\ 
58535.08 & 29.52 & LT/IO:O & $B$ & $18.40\pm0.07$\\ 
58537.17 & 31.61 & LT/IO:O & $B$ & $18.32\pm0.06$\\ 
58541.13 & 35.58 & LT/IO:O & $B$ & $18.46\pm0.03$\\ 
58544.09 & 38.53 & LT/IO:O & $B$ & $18.49\pm0.04$\\ 
58548.95 & 43.40 & LT/IO:O & $B$ & $18.57\pm0.03$\\ 
58554.03 & 48.47 & LT/IO:O & $B$ & $18.74\pm0.02$\\ 
58564.98 & 59.42 & LT/IO:O & $B$ & $19.05\pm0.05$\\ 
58582.05 & 76.49 & LT/IO:O & $B$ & $19.63\pm0.02$\\ 
58582.96 & 77.41 & LT/IO:O & $B$ & $19.59\pm0.04$\\ 
58593.91 & 88.35 & LT/IO:O & $B$ & $19.85\pm0.08$\\ 
58598.95 & 93.39 & LT/IO:O & $B$ & $20.08\pm0.03$\\ 
58602.94 & 97.39 & LT/IO:O & $B$ & $19.92\pm0.07$\\ 
58607.90 & 102.34 & LT/IO:O & $B$ & $20.18\pm0.03$\\ 
58614.03 & 108.47 & LT/IO:O & $B$ & $20.34\pm0.03$\\ 
58625.03 & 119.48 & LT/IO:O & $B$ & $20.48\pm0.04$\\ 
58632.88 & 127.32 & LT/IO:O & $B$ & $20.69\pm0.09$\\ 
58653.90 & 148.34 & LT/IO:O & $B$ & $21.68\pm0.05$\\ 
58659.89 & 154.34 & LT/IO:O & $B$ & $21.63\pm0.06$\\ 
58667.90 & 162.34 & LT/IO:O & $B$ & $22.00\pm0.08$\\ 
\hline
58509.22 & 3.67 & LT/IO:O & $V$ & $19.53\pm0.04$\\ 
58510.19 & 4.64 & LT/IO:O & $V$ & $19.24\pm0.03$\\ 
58511.20 & 5.65 & LT/IO:O & $V$ & $19.01\pm0.02$\\ 
58512.21 & 6.65 & LT/IO:O & $V$ & $18.75\pm0.03$\\ 
58513.23 & 7.68 & LT/IO:O & $V$ & $18.52\pm0.02$\\ 
58514.22 & 8.66 & LT/IO:O & $V$ & $18.30\pm0.02$\\ 
58515.15 & 9.59 & LT/IO:O & $V$ & $18.11\pm0.01$\\ 
58517.15 & 11.60 & LT/IO:O & $V$ & $17.77\pm0.02$\\ 
58518.15 & 12.59 & LT/IO:O & $V$ & $17.65\pm0.04$\\ 
58519.12 & 13.57 & LT/IO:O & $V$ & $17.51\pm0.04$\\ 
58520.18 & 14.63 & LT/IO:O & $V$ & $17.47\pm0.01$\\ 
58521.11 & 15.56 & LT/IO:O & $V$ & $17.41\pm0.01$\\ 
58524.14 & 18.58 & LT/IO:O & $V$ & $17.30\pm0.02$\\ 
58525.07 & 19.52 & LT/IO:O & $V$ & $17.30\pm0.03$\\ 
58528.12 & 22.56 & LT/IO:O & $V$ & $17.25\pm0.04$\\ 
58535.08 & 29.52 & LT/IO:O & $V$ & $17.22\pm0.04$\\ 
58537.17 & 31.62 & LT/IO:O & $V$ & $17.29\pm0.03$\\ 
58541.13 & 35.58 & LT/IO:O & $V$ & $17.34\pm0.02$\\ 
58544.09 & 38.54 & LT/IO:O & $V$ & $17.37\pm0.02$\\ 
58548.96 & 43.40 & LT/IO:O & $V$ & $17.41\pm0.03$\\ 
58554.03 & 48.48 & LT/IO:O & $V$ & $17.54\pm0.01$\\ 
58560.12 & 54.56 & LT/IO:O & $V$ & $17.54\pm0.11$\\ 
58564.98 & 59.43 & LT/IO:O & $V$ & $17.71\pm0.03$\\ 
58582.05 & 76.49 & LT/IO:O & $V$ & $18.06\pm0.03$\\ 
58582.96 & 77.41 & LT/IO:O & $V$ & $18.09\pm0.03$\\ 
58588.88 & 83.33 & LT/IO:O & $V$ & $18.17\pm0.05$\\ 
58593.91 & 88.36 & LT/IO:O & $V$ & $18.31\pm0.04$\\ 
58598.95 & 93.39 & LT/IO:O & $V$ & $18.40\pm0.03$\\ 
58602.95 & 97.39 & LT/IO:O & $V$ & $18.43\pm0.02$\\ 
58607.90 & 102.35 & LT/IO:O & $V$ & $18.55\pm0.03$\\ 
58614.03 & 108.48 & LT/IO:O & $V$ & $18.66\pm0.01$\\ 
58625.03 & 119.47 & LT/IO:O & $V$ & $18.86\pm0.04$\\ 
58632.87 & 127.31 & LT/IO:O & $V$ & $19.02\pm0.04$\\ 
58647.08 & 141.52 & LT/IO:O & $V$ & $19.39\pm0.04$\\ 
58653.89 & 148.33 & LT/IO:O & $V$ & $19.70\pm0.03$\\ 
58659.89 & 154.33 & LT/IO:O & $V$ & $19.78\pm0.02$\\ 
58667.89 & 162.33 & LT/IO:O & $V$ & $20.10\pm0.03$\\ 
58668.89 & 163.33 & LT/IO:O & $V$ & $19.97\pm0.03$\\ 
58676.90 & 171.34 & LT/IO:O & $V$ & $20.27\pm0.03$\\ 
58692.89 & 187.33 & LT/IO:O & $V$ & $20.62\pm0.03$\\ 
58709.88 & 204.33 & LT/IO:O & $V$ & $21.25\pm0.05$\\ 
58710.87 & 205.32 & LT/IO:O & $V$ & $21.48\pm0.06$\\ 
\hline
58509.23 & 3.67 & LT/IO:O & $r'$ & $18.94\pm0.03$\\ 
58510.19 & 4.64 & LT/IO:O & $r'$ & $18.66\pm0.02$\\ 
58511.21 & 5.65 & LT/IO:O & $r'$ & $18.38\pm0.01$\\ 
58512.21 & 6.65 & LT/IO:O & $r'$ & $18.08\pm0.01$\\ 
58513.24 & 7.68 & LT/IO:O & $r'$ & $17.83\pm0.02$\\ 
58514.22 & 8.66 & LT/IO:O & $r'$ & $17.61\pm0.02$\\ 
58515.15 & 9.59 & LT/IO:O & $r'$ & $17.44\pm0.01$\\ 
58517.16 & 11.60 & LT/IO:O & $r'$ & $17.11\pm0.01$\\ 
58518.15 & 12.59 & LT/IO:O & $r'$ & $17.04\pm0.02$\\ 
58519.13 & 13.57 & LT/IO:O & $r'$ & $16.93\pm0.02$\\ 
58520.19 & 14.63 & LT/IO:O & $r'$ & $16.87\pm0.01$\\ 
58521.12 & 15.56 & LT/IO:O & $r'$ & $16.83\pm0.02$\\ 
58524.14 & 18.58 & LT/IO:O & $r'$ & $16.73\pm0.01$\\ 
58525.08 & 19.52 & LT/IO:O & $r'$ & $16.73\pm0.03$\\ 
58528.12 & 22.56 & LT/IO:O & $r'$ & $16.70\pm0.03$\\ 
58535.08 & 29.52 & LT/IO:O & $r'$ & $16.75\pm0.03$\\ 
58537.17 & 31.62 & LT/IO:O & $r'$ & $16.76\pm0.02$\\ 
58541.13 & 35.58 & LT/IO:O & $r'$ & $16.78\pm0.01$\\ 
58544.09 & 38.54 & LT/IO:O & $r'$ & $16.83\pm0.01$\\ 
58548.96 & 43.40 & LT/IO:O & $r'$ & $16.91\pm0.02$\\ 
58554.03 & 48.48 & LT/IO:O & $r'$ & $16.97\pm0.01$\\ 
58560.12 & 54.57 & LT/IO:O & $r'$ & $17.06\pm0.05$\\ 
58564.98 & 59.43 & LT/IO:O & $r'$ & $17.12\pm0.02$\\ 
58582.05 & 76.50 & LT/IO:O & $r'$ & $17.36\pm0.01$\\ 
58582.97 & 77.41 & LT/IO:O & $r'$ & $17.38\pm0.01$\\ 
58588.89 & 83.33 & LT/IO:O & $r'$ & $17.46\pm0.02$\\ 
58593.91 & 88.36 & LT/IO:O & $r'$ & $17.52\pm0.02$\\ 
58598.95 & 93.40 & LT/IO:O & $r'$ & $17.59\pm0.02$\\ 
58602.95 & 97.39 & LT/IO:O & $r'$ & $17.63\pm0.01$\\ 
58607.90 & 102.35 & LT/IO:O & $r'$ & $17.72\pm0.01$\\ 
58614.04 & 108.48 & LT/IO:O & $r'$ & $17.77\pm0.02$\\ 
58625.03 & 119.47 & LT/IO:O & $r'$ & $17.95\pm0.02$\\ 
58632.87 & 127.31 & LT/IO:O & $r'$ & $18.03\pm0.03$\\ 
58647.08 & 141.52 & LT/IO:O & $r'$ & $18.36\pm0.03$\\ 
58653.89 & 148.33 & LT/IO:O & $r'$ & $18.57\pm0.05$\\ 
58659.89 & 154.33 & LT/IO:O & $r'$ & $18.72\pm0.01$\\ 
58667.89 & 162.33 & LT/IO:O & $r'$ & $18.95\pm0.02$\\ 
58668.88 & 163.33 & LT/IO:O & $r'$ & $18.93\pm0.02$\\ 
58676.89 & 171.34 & LT/IO:O & $r'$ & $19.07\pm0.02$\\ 
58692.88 & 187.33 & LT/IO:O & $r'$ & $19.45\pm0.03$\\ 
58709.87 & 204.32 & LT/IO:O & $r'$ & $20.01\pm0.03$\\ 
58710.87 & 205.31 & LT/IO:O & $r'$ & $19.98\pm0.03$\\ 
\hline
58509.23 & 3.67 & LT/IO:O & $i'$ & $18.24\pm0.02$\\ 
58510.20 & 4.64 & LT/IO:O & $i'$ & $17.99\pm0.02$\\ 
58511.21 & 5.65 & LT/IO:O & $i'$ & $17.64\pm0.02$\\ 
58512.21 & 6.66 & LT/IO:O & $i'$ & $17.36\pm0.01$\\ 
58513.24 & 7.68 & LT/IO:O & $i'$ & $17.10\pm0.02$\\ 
58514.22 & 8.67 & LT/IO:O & $i'$ & $16.90\pm0.01$\\ 
58515.15 & 9.59 & LT/IO:O & $i'$ & $16.77\pm0.02$\\ 
58517.16 & 11.60 & LT/IO:O & $i'$ & $16.50\pm0.01$\\ 
58518.15 & 12.59 & LT/IO:O & $i'$ & $16.42\pm0.01$\\ 
58519.13 & 13.57 & LT/IO:O & $i'$ & $16.37\pm0.02$\\ 
58520.19 & 14.63 & LT/IO:O & $i'$ & $16.34\pm0.03$\\ 
58521.12 & 15.56 & LT/IO:O & $i'$ & $16.29\pm0.01$\\ 
58524.14 & 18.58 & LT/IO:O & $i'$ & $16.23\pm0.01$\\ 
58525.08 & 19.52 & LT/IO:O & $i'$ & $16.20\pm0.04$\\ 
58528.12 & 22.56 & LT/IO:O & $i'$ & $16.18\pm0.04$\\ 
58535.08 & 29.53 & LT/IO:O & $i'$ & $16.23\pm0.02$\\ 
58537.17 & 31.62 & LT/IO:O & $i'$ & $16.26\pm0.03$\\ 
58541.14 & 35.58 & LT/IO:O & $i'$ & $16.29\pm0.01$\\ 
58544.09 & 38.54 & LT/IO:O & $i'$ & $16.32\pm0.01$\\ 
58548.96 & 43.40 & LT/IO:O & $i'$ & $16.38\pm0.02$\\ 
58554.03 & 48.48 & LT/IO:O & $i'$ & $16.43\pm0.01$\\ 
58560.12 & 54.57 & LT/IO:O & $i'$ & $16.50\pm0.05$\\ 
58564.98 & 59.43 & LT/IO:O & $i'$ & $16.54\pm0.02$\\ 
58582.05 & 76.50 & LT/IO:O & $i'$ & $16.70\pm0.01$\\ 
58582.97 & 77.41 & LT/IO:O & $i'$ & $16.69\pm0.02$\\ 
58588.89 & 83.33 & LT/IO:O & $i'$ & $16.77\pm0.01$\\ 
58593.91 & 88.36 & LT/IO:O & $i'$ & $16.80\pm0.02$\\ 
58598.95 & 93.40 & LT/IO:O & $i'$ & $16.86\pm0.02$\\ 
58602.95 & 97.39 & LT/IO:O & $i'$ & $16.92\pm0.02$\\ 
58607.90 & 102.35 & LT/IO:O & $i'$ & $16.97\pm0.01$\\ 
58614.04 & 108.48 & LT/IO:O & $i'$ & $17.04\pm0.02$\\ 
58625.02 & 119.47 & LT/IO:O & $i'$ & $17.16\pm0.01$\\ 
58632.86 & 127.31 & LT/IO:O & $i'$ & $17.30\pm0.04$\\ 
58647.07 & 141.52 & LT/IO:O & $i'$ & $17.53\pm0.03$\\ 
58653.88 & 148.33 & LT/IO:O & $i'$ & $17.70\pm0.02$\\ 
58659.88 & 154.33 & LT/IO:O & $i'$ & $17.81\pm0.03$\\ 
58667.88 & 162.33 & LT/IO:O & $i'$ & $17.91\pm0.02$\\ 
58668.88 & 163.33 & LT/IO:O & $i'$ & $17.94\pm0.02$\\ 
58676.89 & 171.33 & LT/IO:O & $i'$ & $18.11\pm0.01$\\ 
58692.88 & 187.33 & LT/IO:O & $i'$ & $18.40\pm0.02$\\ 
58709.87 & 204.32 & LT/IO:O & $i'$ & $18.89\pm0.02$\\ 
58710.86 & 205.31 & LT/IO:O & $i'$ & $19.02\pm0.02$\\ 
\hline
58513.28 & 7.73 & LT/IO:O & $z'$ & $16.67\pm0.03$\\ 
58514.22 & 8.67 & LT/IO:O & $z'$ & $16.54\pm0.02$\\ 
58515.15 & 9.59 & LT/IO:O & $z'$ & $16.38\pm0.03$\\ 
58517.16 & 11.60 & LT/IO:O & $z'$ & $16.22\pm0.04$\\ 
58518.15 & 12.59 & LT/IO:O & $z'$ & $16.16\pm0.02$\\ 
58519.13 & 13.57 & LT/IO:O & $z'$ & $16.13\pm0.06$\\ 
58520.19 & 14.63 & LT/IO:O & $z'$ & $16.09\pm0.02$\\ 
58521.12 & 15.56 & LT/IO:O & $z'$ & $16.03\pm0.02$\\ 
58524.14 & 18.58 & LT/IO:O & $z'$ & $16.02\pm0.02$\\ 
58525.08 & 19.52 & LT/IO:O & $z'$ & $15.96\pm0.06$\\ 
58528.12 & 22.56 & LT/IO:O & $z'$ & $15.97\pm0.02$\\ 
58535.08 & 29.53 & LT/IO:O & $z'$ & $16.02\pm0.03$\\ 
58537.17 & 31.62 & LT/IO:O & $z'$ & $16.04\pm0.05$\\ 
58541.14 & 35.58 & LT/IO:O & $z'$ & $16.05\pm0.02$\\ 
58544.09 & 38.54 & LT/IO:O & $z'$ & $16.07\pm0.03$\\ 
58548.96 & 43.40 & LT/IO:O & $z'$ & $16.09\pm0.02$\\ 
58554.03 & 48.48 & LT/IO:O & $z'$ & $16.15\pm0.02$\\ 
58560.12 & 54.57 & LT/IO:O & $z'$ & $16.14\pm0.07$\\ 
58564.98 & 59.43 & LT/IO:O & $z'$ & $16.14\pm0.04$\\ 
58582.05 & 76.50 & LT/IO:O & $z'$ & $16.35\pm0.01$\\ 
58582.97 & 77.41 & LT/IO:O & $z'$ & $16.36\pm0.02$\\ 
58588.89 & 83.33 & LT/IO:O & $z'$ & $16.40\pm0.04$\\ 
58593.91 & 88.36 & LT/IO:O & $z'$ & $16.43\pm0.03$\\ 
58598.95 & 93.40 & LT/IO:O & $z'$ & $16.48\pm0.02$\\ 
58602.95 & 97.39 & LT/IO:O & $z'$ & $16.52\pm0.04$\\ 
58607.90 & 102.35 & LT/IO:O & $z'$ & $16.58\pm0.02$\\ 
58614.04 & 108.48 & LT/IO:O & $z'$ & $16.62\pm0.02$\\ 
58625.02 & 119.47 & LT/IO:O & $z'$ & $16.73\pm0.01$\\ 
58632.86 & 127.31 & LT/IO:O & $z'$ & $16.84\pm0.06$\\ 
58647.07 & 141.52 & LT/IO:O & $z'$ & $17.03\pm0.01$\\ 
58653.88 & 148.33 & LT/IO:O & $z'$ & $17.15\pm0.02$\\ 
58659.88 & 154.32 & LT/IO:O & $z'$ & $17.24\pm0.01$\\ 
58667.88 & 162.33 & LT/IO:O & $z'$ & $17.37\pm0.01$\\ 
58668.88 & 163.32 & LT/IO:O & $z'$ & $17.37\pm0.02$\\ 
58676.89 & 171.33 & LT/IO:O & $z'$ & $17.48\pm0.02$\\ 
58692.88 & 187.32 & LT/IO:O & $z'$ & $17.71\pm0.02$\\ 
58709.87 & 204.31 & LT/IO:O & $z'$ & $18.08\pm0.03$\\ 
58710.86 & 205.31 & LT/IO:O & $z'$ & $18.12\pm0.06$\\ 
\hline
58585.92 & 80.37 & GTC/EMIR & $J$ & $16.08\pm0.05$\\ 
58685.87 & 180.32 & NOT/NOTCam & $J$ & $16.98\pm0.05$\\ 
\hline
58515.13 & 9.58 & LT/IO:I & $H$ & $15.94\pm0.17$\\ 
58517.14 & 11.58 & LT/IO:I & $H$ & $15.90\pm0.04$\\ 
58520.19 & 14.64 & LT/IO:I & $H$ & $15.81\pm0.04$\\ 
58524.09 & 18.53 & LT/IO:I & $H$ & $15.78\pm0.05$\\ 
58528.12 & 22.57 & LT/IO:I & $H$ & $15.71\pm0.05$\\ 
58537.15 & 31.59 & LT/IO:I & $H$ & $15.79\pm0.06$\\ 
58541.12 & 35.56 & LT/IO:I & $H$ & $15.88\pm0.05$\\ 
58548.96 & 43.41 & LT/IO:I & $H$ & $15.94\pm0.05$\\ 
58560.13 & 54.57 & LT/IO:I & $H$ & $16.04\pm0.04$\\ 
58582.06 & 76.50 & LT/IO:I & $H$ & $16.10\pm0.05$\\ 
58582.97 & 77.41 & LT/IO:I & $H$ & $16.14\pm0.04$\\ 
58585.93 & 80.37 & GTC/EMIR & $H$ & $16.13\pm0.04$\\ 
58593.98 & 88.43 & LT/IO:I & $H$ & $16.18\pm0.06$\\ 
58602.95 & 97.40 & LT/IO:I & $H$ & $16.24\pm0.05$\\ 
58614.04 & 108.49 & LT/IO:I & $H$ & $16.36\pm0.06$\\ 
58625.04 & 119.48 & LT/IO:I & $H$ & $16.35\pm0.05$\\ 
58647.09 & 141.53 & LT/IO:I & $H$ & $16.40\pm0.06$\\ 
58659.90 & 154.34 & LT/IO:I & $H$ & $16.59\pm0.04$\\ 
58676.90 & 171.35 & LT/IO:I & $H$ & $16.65\pm0.04$\\ 
58685.89 & 180.33 & NOT/NOTCam & $H$ & $16.80\pm0.06$\\ 
58692.90 & 187.34 & LT/IO:I & $H$ & $16.81\pm0.04$\\ 
58709.89 & 204.33 & LT/IO:I & $H$ & $16.95\pm0.06$\\ 
58710.88 & 205.32 & LT/IO:I & $H$ & $16.95\pm0.06$\\ 
\hline
58585.93 & 80.38 & GTC/EMIR & $K_{\mathrm{s}}$ & $16.25\pm0.07$\\ 
58647.93 & 142.37 & NOT/NOTCam & $K_{\mathrm{s}}$ & $16.72\pm0.07$\\ 
58685.90 & 180.34 & NOT/NOTCam & $K_{\mathrm{s}}$ & $16.91\pm0.06$\\ 
\hline\hline
\end{supertabular}
\end{appendix}


\end{document}